\documentclass{aa}  

\usepackage{graphicx}
\usepackage{txfonts}
\begin{document}

   \title{Galaxy morphoto-Z with neural Networks (GaZNets). I. Optimized accuracy and outlier fraction from Imaging and Photometry}
   
   \authorrunning{Li et al.}
   \titlerunning{Galaxy morphoto-$z$ with Neural Networks}

   \author{Rui Li\inst{1,2},
          Nicola R. Napolitano\inst{3,4},
          Haicheng Feng\inst{1,5},
          Ran Li\inst{1,2},
          Valeria Amaro\inst{3},
          Linghua Xie\inst{3},
          Crescenzo Tortora\inst{6},
          Maciej Bilicki\inst{7},
          Massimo Brescia\inst{6},
          Stefano Cavuoti\inst{6,8},
          \and
          Mario Radovich\inst{9}
          }

   \institute{
     School of Astronomy and Space Science, University of Chinese Academy of Sciences, Beijing 100049, China\\
     \email{liruiww@gmail.com}
     \and
     National Astronomical Observatories, Chinese Academy of Sciences, 20A Datun Road, Chaoyang District, Beijing 100012, China
     \and 
     School of Physics and Astronomy, Sun Yat-sen University, Zhuhai Campus, 2 Daxue Road, Xiangzhou District, Zhuhai, P. R. China\\ \email{napolitano@mail.sysu.edu.cn}
     \and
     CSST Science Center for Guangdong-Hong Kong-Macau Great Bay Area, Zhuhai, China, 519082
     \and
     Yunnan Observatories, Chinese Academy of Sciences, Kunming, 650011, Yunnan, People's Republic of China
     \and INAF -- Osservatorio Astronomico di Capodimonte, Salita Moiariello 16, 80131 - Napoli, Italy
     \and
     Center for Theoretical Physics, Polish Academy of Sciences, Al. Lotnik{\'o}w 32/46, 02-668 Warsaw, Poland
     \and 
     INFN - Sezione di Napoli, via Cinthia 9, 80126 Napoli, Italy
     \and
     INAF - Osservatorio Astronomico di Padova, via dell'Osservatorio 5, 35122 Padova, Italy
             }

   \date{Received September 15, 1996; accepted March 16, 1997}

% \abstract{}{}{}{}{} 
% 5 {} token are mandatory
 
  \abstract
{}
{In the era of large sky surveys, photometric redshifts (photo-$z$) represent crucial information for
galaxy evolution and cosmology studies. In this work, we propose a new Machine Learning (ML) tool called Galaxy morphoto-Z with neural Networks (GaZNet-1), which uses both images and multi-band photometry measurements to predict galaxy redshifts, with accuracy, precision and outlier fraction superior to standard methods based on photometry only.}
{As a first application of this tool, we estimate photo-$z$ of a sample of galaxies in the Kilo-Degree Survey (KiDS). GaZNet-1 is trained and tested on $\sim140\,000$ galaxies collected from KiDS Data Release 4 (DR4), for which spectroscopic redshifts are available from different surveys. 
This sample is dominated by bright (MAG$\_$AUTO$<21$) and low redshift ($z<0.8$) systems, however, we could use $\sim 6500$ galaxies in the range $0.8<z<3$
%can be used for extending 
to effectively extend the training to higher redshift. The inputs are the $r$-band galaxy images plus the 9-band magnitudes and colors, from the combined catalogs of optical photometry from KiDS and near-infrared photometry from the VISTA Kilo-degree Infrared survey.}
{By combining the images and catalogs, GaZNet-1 can achieve extremely high precision in normalized  median  absolute  deviation (NMAD=0.014 for lower redshift and NMAD=0.041 for higher redshift galaxies) and low fraction of outliers ($0.4\%$ for lower and $1.27\%$ for higher redshift galaxies).
Compared to ML codes
%code we have constructed following the same architecture, but 
using only photometry as input, GaZNet-1 also shows a $\sim 10-35$\% improvement in precision at different redshifts and a $\sim 45$\% reduction in the fraction of outliers. We finally discuss that, by correctly separating galaxies from stars and active galactic nuclei, the overall photo-$z$ outlier fraction of galaxies can be cut down to 0.3\%.}
{}


\maketitle
%
%-------------------------------------------------------------------

\section{Introduction}
\label{sec:intro}
In the last decade, the Stage  \uppercase\expandafter{\romannumeral 3}
% third generation 
sky surveys, e.g., Kilo-Degree Survey (KiDS, \citealt{deJong2013ExA....35...25D}), Hyper Suprime-Cam (HSC; \citealt{Aihara+18_HSC}), Dark Energy Survey (DES; \citealt{2005+DES}), have provided images of hundreds of millions of galaxies at optical or near-infrared (NIR) wavelengths. These surveys have achieved significant advances in cosmology (e.g., \citealt{Hildebrandt2017MNRAS.465.1454H}; \citealt{Hikage2019PASJ...71...43H}; \citealt{Abbott2022PhRvD.105b3520A}) and galaxy formation and evolution (e.g., \citealt{Roy+18}; \citealt{Greco2018ApJ...857..104G}; \citealt{Goulding2018PASJ...70S..37G}; \citealt{Adhikari2021ApJ...923...37A}), but, at the same time, have left many open questions about the overall cosmological model (\citealt{2021APh...13102606D}).

In the next decade, the Stage \uppercase\expandafter{\romannumeral 4}
%fourth generation 
surveys (\citealt{Weinberg2013PhR...530...87W}) , e.g., Euclid (\citealt{Laureijs+11_Euclid}), Vera Rubin Legacy Survey in Space and Time (VR/LSST; \citealt{Izevic+19_LSST}), China Space Station Telescope (CSST; \citealt{Zhan+18_csst}), will observe billions of galaxies with photometric bands ranging from the ultraviolet to the NIR. 
This unprecedented amount of
%galaxies, people will get much deeper insight into many 
data will help us to get a deeper insight into cosmology and galaxy evolution.
For instance, we will be able to gain a more detailed understanding of the dark matter distribution in the universe, constrain the equation of state of the dark energy
%cosmology parameters constrained 
with weak lensing (e.g., \citealt{Laureijs2011arXiv1110.3193L}; \citealt{Hildebrandt2017MNRAS.465.1454H};
\citealt{Abbott2018PhRvD..98d3526A};
\citealt{Gong2019ApJ...883..203G};
\citealt{Joachimi2021A&A...646A.129J};
\citealt{Heymans2021A&A...646A.140H}), study the mass-size relation of galaxies at higher redshift ($z>1.0$), and explore the stellar and dark matter assembly in galaxies and clusters (e.g., \citealt{Yang2012ApJ...752...41Y}; \citealt{Moster2013MNRAS.428.3121M}; \citealt{Behroozi2019MNRAS.488.3143B}, \citealt{2022FrASS...8..197N}) over enormous statistical samples.

%In the next decade, the fourth generation surveys, e.g., Euclid (\citealt{Laureijs+11_Euclid}), Large Synoptic Survey Telescope (LSST; \citealt{Izevic+19_LSST}), China Space Station Telescope (CSST; \citealt{Zhan+18_csst}), will observe billions of galaxies with photometric bands ranging from the ultraviolet to the NIR. This unprecedented amount of data will help us to get a deeper insight into
%galaxy evolution and cosmology. For instance, we will be able to gain a more detailed understanding of the mass-size relation of galaxies at higher redshift ($z>1.0$), explore the stellar and dark matter
%assembly in galaxies and clusters (e.g., \citealt{Yang2012ApJ...752...41Y}; \citealt{Moster2013MNRAS.428.3121M}; \citealt{Behroozi2019MNRAS.488.3143B}, \citealt{2022FrASS...8..197N}) over enormous statistical samples, study the dark matter distribution in the universe, and constrain the equation of state of the dark energy
%with weak lensing (e.g., \citealt{Laureijs2011arXiv1110.3193L}; \citealt{Hildebrandt2017MNRAS.465.1454H};
%\citealt{Abbott2018PhRvD..98d3526A};
%\citealt{Gong2019ApJ...883..203G};
%\citealt{Joachimi2021A&A...646A.129J};
%\citealt{Heymans2021A&A...646A.140H}). 

To achieve real breakthroughs in these areas,
accurate galaxy redshifts are essential, as, by providing object distances and lookback time, they permit to trace those objects back in time. 
Precise redshifts can only be estimated from galaxy spectra: current spectroscopic surveys, such as the Sloan Digital Sky Survey (SDSS, \citealt{Ahumada2020ApJS..249....3A}), the Galaxy and Mass Assembly (GAMA \citealt{Baldry2018MNRAS.474.3875B}) and the Dark Energy Spectroscopic Instrument (DESI, \citealt{DESI_Collaboration_2016}) have collected data for millions of galaxies, while future surveys, e,g, the 4-meter Multi-Object Spectroscopic Telescope (4MOST, \citealt{deJong2019_4MOST}), plan to expand spectroscopic measurements to samples of
hundreds of millions of galaxies.
However, due to the limited observation depth and prohibitive exposure times, it is impossible to spectroscopically follow up the even larger and fainter samples of billions of galaxies expected in future imaging surveys.

A fast, low-cost alternative is offered by 
%deriving the 
photometric redshifts (photo-$z$) estimated from deep, multi-band photometry.
The idea of photo-$z$ was initially proposed by \cite{1962IAUS...15..390B}, where they used a redshift-magnitude relation to predict the redshifts from the galaxy luminosities.
%measured the redshifts of elliptical galaxies in distant clusters through a redshift-magnitude relation. 
Without spectroscopic observations and the knowledge of galaxy evolution, the relation could still provide acceptable redshifts, even using only a limited number of filters.
%the filters are not too many. 
Later, this method was adopted to 
%many works to 
extensively estimate galaxy redshifts (e.g., \citealt{Couch1983, Koo1985AJ, Connolly1995AJ, Connolly1997hst,Wang1998AJ}). However, 
%there is a shortness that can not be ignored
albeit straightforward, this method has some limitations: 1) the redshift-magnitude relation 
%needs to be 
is inferred in advance from bright galaxies
%for which the redshift have been \hai{derived} from high quality spectra, 
via spectroscopy, and 2) the relation is hard to extend to fainter galaxies.
Another method used to determine photo-$z$ is spectral-energy-distribution (SED, hereafter) fitting. 
This method is based on galaxy templates, both theoretical and empirical.

\begin{figure*}
\centerline{\includegraphics[width=15.5cm]{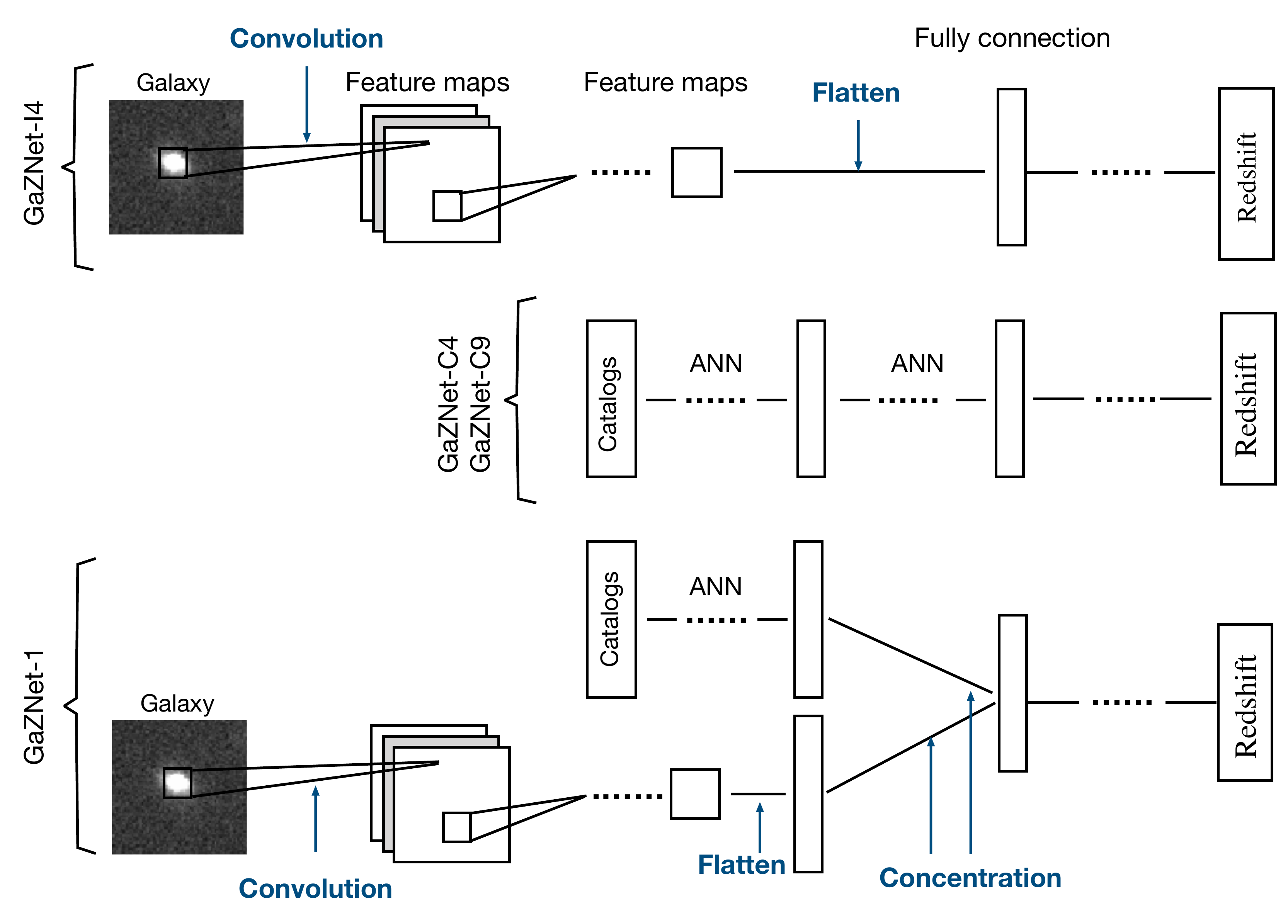}}
\caption{The Machine Learning models are used in this work.  Up: CNN structure of GaZNet-I4 with only galaxy images as input.  Middle:  ANN structure of GaZNet-C4 and GaZNet-C9, with only catalog as input;  bottom: structure of GaZNet-1, fed by both galaxies images and the corresponding catalogs.}
\label{fig:CNN_model}
\vspace{0.5cm}
\end{figure*}

%#### From here

By fitting the observed multi-band photometry to the SED from galaxy templates, one can infer individual galaxy photo-$z$. With knowledge of galaxy types and their evolution with redshift, this method can be expanded to faint galaxies, and even extrapolated to redshifts higher than the spectroscopic limit.
There is a variety of 
%firstly publicly available 
photo-$z$ codes based on SED fitting. Among the most popular ones, there is HyperZ (\citealt{Bolzonella2000A&A}),
%. This software 
%Given the 
which makes use of multi-band magnitudes of galaxies and the corresponding errors
to best fit the SED templates by minimizing a given $\chi^2$ function. 
%Two kinds of SED templates are used in HyperZ: CWW templates (\citealt{Coleman1980ApJS_CWW}) and \hai{theoretically synthetic templates (BC03; \citealt{BC03})}. The reddening laws and the damping of the Ly$\alpha$ forest also have been considered in the fitting process. 
An extension of HyperZ, known as Bayesian photometric redshifts (BPZ, \citealt{2000ApJ...536..571B_BPZ}), is another popular photo-$z$ tool. Instead of simple 
%minimizing a given 
$\chi^2$ minimization, BPZ introduces the prior knowledge of the redshift distribution of magnitude-limited samples under a
%magnitudes under the 
Bayesian framework, which effectively reduces the number of catastrophic outliers in the predictions.
%Apart from the CWW templates, two more templates of star-bursting galaxies have been added to BPZ. 

%Apart from the reshift-magnitude relation and the SED fitting, 
Besides these fitting tools, Machine Learning (ML) algorithms, especially Artificial Neural Networks (ANNs), have started to be extensively used to determine galaxy photo-$z$ (e.g., \citealt{Collister2007MNRAS, Abdalla2008MNRAS, Banerji2008MNRAS}). Given a training sample of galaxies with spectroscopic redshifts, ML algorithms can learn the relationship between redshift and multi-band photometry. If the training sample covers a representative redshift range and the ML model is well trained, photo-$z$ can be obtained with extremely high precision. Different tools for photo-$z$ based on ML have been successfully tested on multi-band photometry data, for example estimating photo-$z$ with ANNs (ANNz, \citealt{Collister2004PASP..116..345C}; ANNz2 \citealt{Sadeh2016PASP..128j4502S}) or the Multi-Layer Perceptron trained with Quasi-Newton Algorithm (MLPQNA, \citealt{Cavuoti2012A&A_MLPQNA}, \citealt{Amaro2021+photz}).

%MLPQNA is an effective computing implementation of neural networks exploited for the first time to solve regression problems in the astrophysical context.

%A particularly promising tool for photo-$z$ is the Multi-Layer Perceptron trained with Quasi-Newton Algorithm (MLPQNA, \citealt{Cavuoti2012A&A_MLPQNA}). MLPQNA is an effective computing implementation of neural networks exploited for the first time to solve regression problems in the astrophysical context. A test on PHAT1 dataset (\citealt{Hildebrandt2010A&A}) indicated that, MLPQNA, with smaller bias and less outliers, performs better than most of the traditional standard SED fitting methods. This code have been used in some of the current sky surveys, e.g. KiDS (\citealt{Cavuoti+15_KIDS_I}) and \hai{Sloan Digital Sky Survey (SDSS; }\citealt{Brescia2014A&A}).

%However, the performance of the ML methods strongly rely on the training data. 
%Evidences have shown that 
%For instance, ML method can predict more accurate photo-$z$ than SED fitting at lower redshift (e.g. $z<1$), because enough high quality galaxies can be used to train the ML (\citealt{Abdalla2008MNRAS}). 
%Going to higher redshift, where the number of the training galaxies sharply drop and the quality of the photometry data also decreases, the errors of the predictions will become larger than SED fitting.

{
%All of the methods mentioned above are based on the catalog of multi-band photometry.  
%being based on mapping between aperture photometry and spectroscopic redshifts,
%Therefore, 
Accurate photometry measurements are extremely important for ML and SED fitting methods, as
the presence of noisy or biased photometry 
%or if the measurement is not accuracy, 
would end up in large scatter and a high outlier fraction in the predicted values.
%would occurs.
%although ML are less sensitive than SED fitting methods to calibration problems which might affect the combinations of datasets coming from different programs (reference?).
%(and also for SED fitting) to make predictions on photo-$z$. 
For instance, biased photometry is typically produced in the case of close galaxy pairs or in the presence of bright neighbors. 
On top of that, there are well-known degeneracies between colors and redshift plaguing late-type systems, in particular, as high-$z$ star-forming galaxies can be confused with lower redshift ellipticals. 

These examples suggest that there might be some crucial information encoded in images that can help solving typical systematics, affecting the methods based on photometry only.}

%\hai{The SEDs of different type galaxies might degenerate with each other in various redshifts. Moreover, }some kind of galaxies, e.g. spirals, pairs or the ones with brighter neighbours, correct magnitudes are hard to be got, which could lead biases on the photo-$z$ determinations. 

%Therefore, apart from the photometric catalogs,
%of \hai{photometry}, 
%if the images themselves can be provided, worse predictions \hai{caused by degeneracy of SED and wrong magnitude} could be corrected. 
ML has been shown to be able to learn
%On the other hand, from images, ML can learn
%, for example, 
galaxy properties like their size, morphology, and their environment from images. This information can help suppress catastrophic errors and improve the accuracy of the photo-$z$ predictions. In recent years, many studies have been
%some works that 
trying to estimate photo-$z$ directly from multi-band images using deep learning. A first attempt was presented by \citet{Hoyle2016A&C....16...34H}, 
%in which a new method to 
where they estimated photo-$z$  
%passing the full galaxy imaging into 
with a Deep Neural Network (DNN) applied to full galaxy imaging data. Lately, a similar approach has been applied to data from the Sloan Digital Sky Survey and Hyper Suprime-Cam Subaru Strategic Program (e.g., \citealt{D'Isanto2018A&A...609A.111D, Pasquet2019A&A...621A..26P, Schuldt2021A&A...651A..55S, Dey2021arXiv211203939D}). These analyses showed that unbiased photo-$z$ can be estimated directly from multi-band images.
A more simplistic method for taking morphology features into account - e.g. size, ellipticity and S{\'e}rsic index - has been proposed in \citet{Soo2018MNRAS.475.3613S}, where they added structural parameters to the photometric catalogs used in standard ANNs.

In this paper, we develop a new ML method to estimate the {\it morphoto-$z$}, i.e. redshifts estimated from the combination of images and catalogs of photometry and color measurements. In the following, we distinguish these morphoto-$z$ from the redshift predicted from photometry only, the classical {\it photo-$z$}, and from the ones obtained from images only, which - for convenience - we call {\it morpho-$z$}.
%\footnote{\rui{
This is the first time such a technique has been developed and applied to real data: specifically, we will use optical images and optical+NIR multi-band photometry from the KiDS survey. Just before the submission of this paper, a similar approach was proposed by \citet{Zhou2021arXiv211208690Z}, but this latter work is based on (CSST) simulated data only.
%work has been done on CSST simulation data, in which they claim that by combining the catalog and images, one can achieve higher accuracy in the photo-$z$ determination (\citealt{Zhou2021arXiv211208690Z}).
%}}. 

This work is organized as follows. In \S2, we describe how to build the ML models and to collect the training and testing samples. In \S3, we train the networks and show the performance of the tools. In \S4 and \S5, we discuss the results and draw some conclusions.

%--------------------------------------------------------------------
\section{The ML method}
In this work, we intend to couple standard ML regression tools, usually applied to galaxy multi-band photometry, with Deep Learning techniques, to improve the estimates of galaxy redshifts using the information from features distilled from galaxy images. 

In particular, we want to address the following questions:
1) can redshifts be estimated directly from multi-band images of KiDS galaxies, and how does typical accuracy compare to ML tools based on integrated photometry and color measurements only? 2) If any, how much improvement in precision and scatter can images and catalogs add to tools combining all together? 

To answer these questions, we %compare the performances of 
have developed and compared four ML tools 
%that we have developed 
to estimate the galaxy redshifts. These differ - among each other - for the type of input data they can work with.
%according to the different inputs, and compare their performance. 
In this section, we start by describing the structures and the training of these four tools.

\subsection{Network architectures}
\label{sec:arch}
The ML parts of our networks are constituted by ANNs.
%A commonly used ML tool to determine galaxy photo-$z$s 
%for galaxies is 
%the ANNs, %and it have 
These have been proved to work well on catalogs made of magnitudes and color measurements (e.g., \citealt{Collister2007MNRAS, Abdalla2008MNRAS, Banerji2008MNRAS,Cavuoti+15_KIDS_I, Brescia2014A&A,deJong2017A&A...604A.134D, Bilicki2018A&A...616A..69B,Bilicki2021A&A...653A..82B}). A typical ANN structure consists of three main parts: input, hidden, and output layers. The input and the output layers are used to load the data in the network and to issue the predictions. The hidden layers, composed of fully connected artificial neurons in a sequence of multiple layers, are used to extract features. These features are subsequently abstracted to allow the networks to determine the final outputs.
In redshift estimates, the inputs of the networks are catalogs of some form of multi-band aperture photometry of galaxies, i.e. a measurements of the total flux in different %wavelength ranges
filters , usually from optical to the NIR wavelengths. 
%\rui{Occasionally, other morphology features, e.g. size, ellipticity, S{\'e}rsic index, are also added to the catalogs used as ANN inputs (e.g., \citealt{Soo2018MNRAS.475.3613S}).}

The Deep Learning components of the four tools are constituted by Convolutional Neural Networks (CNN,  \citealt{1990Handwritten}), which are an effective family of
algorithms for feature extraction from images. 
CNNs mimic the biological perception mechanisms with convolution operations.
%CNNs make use of convolution operations to mimic the biological perception mechanisms. 
%with convolution calculations, 
This makes them specially suitable for image processing, pattern recognition and other tasks relative to images (e.g., \citealt{2018MNRAS.476.3661D, 2018MNRAS.479..415A,2020MNRAS.491.1554W,2020A&C....3200390C}; \citealt{Li2020_DR4lens,Li2021_GaLNet,Li2021_DR5lens}; \citealt{2021ApJ...916....4T}). The CNNs have become popular years after its introduction, because of the significant progress in the graphics processing unit (GPU) technology.
%in computer science. 

\begin{figure*}
\centerline{\includegraphics[width=18cm]{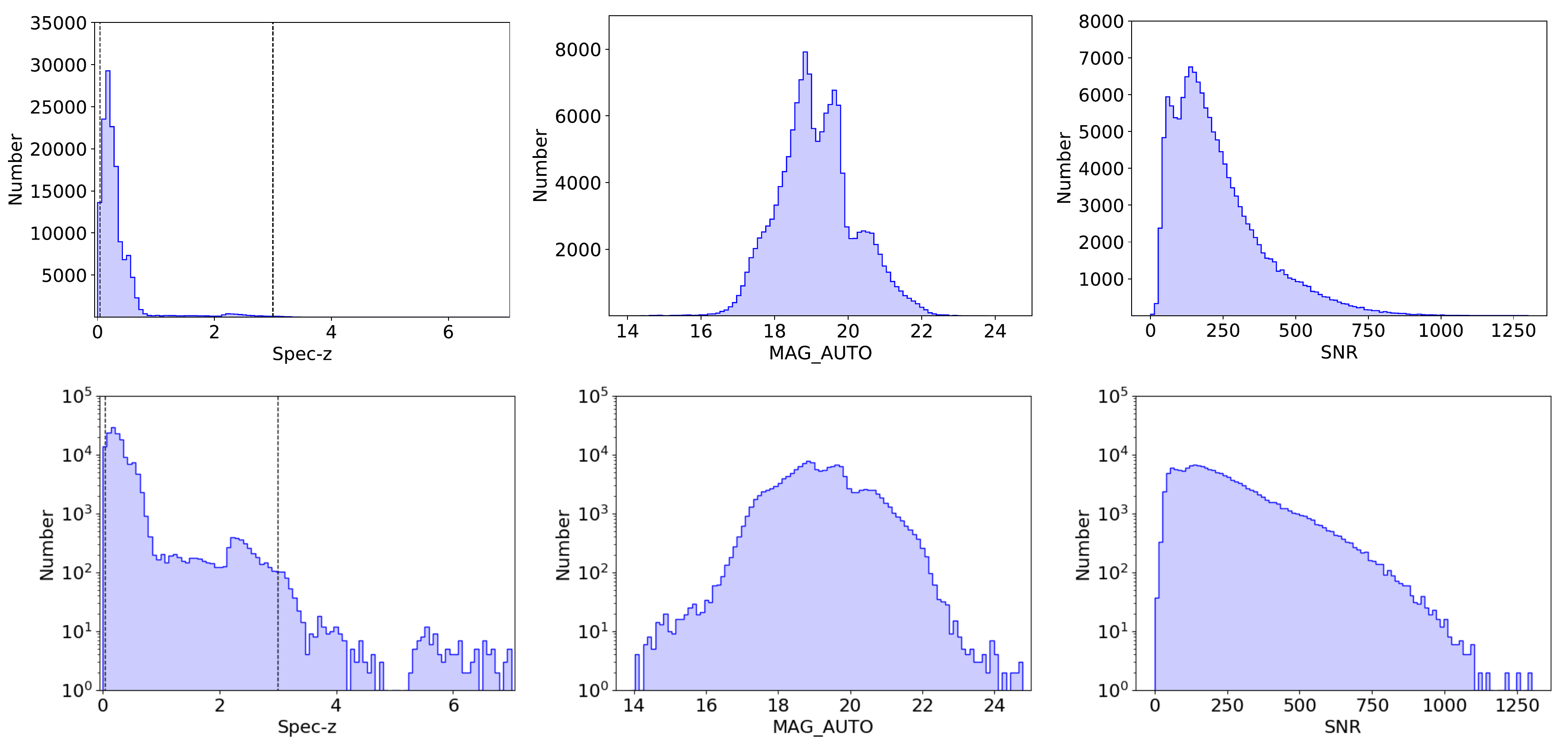}}
\caption{The distribution of some relevant parameters of the training and testing data. In the top row, number counts are in linear scale, while in the bottom row number counts are are in logarithmic scale. The first panel on the left shows the original spectroscopic sample of the 148\,521 galaxies collected in KiDS+VIKING. However, only the 134\,148 galaxies located between the two vertical dashed lines (spec-$z$$=0.04$ and spec-$z$$=3$) are used in this work for training and testing the GaZNets. For these galaxies, we show the MAG$\_$AUTO and SNRs in the second and third panels.}
\label{fig:para_distri}
%\vspace{1cm}
\end{figure*}

Here below, we introduce the structures of the first series of algorithms for ``Galaxy morphoto-Z with neural Networks'' (GaZNets, hereafter). These are introduced to perform galaxy morpho/photo-$z$s using different combinations of inputs, including multi-band photometry and imaging (see Fig. \ref{fig:CNN_model}). In details:
%In this work, the GaZNet, as well as the other three tools, GaZNet$-$I,GaZNet$-$C4 and GaZNet$-$C9 (as shown in Fig. \ref{fig:CNN_model})  are disigned as bellow:
\begin{itemize}
\item {\bf GaZNet$-$I4}. This is a CNN model, which makes use of 4 optical bands ($u, g, r, i$) galaxy images, with a cutout size of $8\times8''$ (corresponding to $40\times40$ pixels, see \S\ref{sec:train_test_data}), as input. The model is a slightly modified architecture from VGGNet (\citealt{Simonyan2014}). It is constituted of four blocks made of different numbers of convolutional layers. Each of the first two blocks contains two layers, and of the other two blocks contain three layers. 
After the four blocks, a flatten layer is used to transform the high dimensional features into one-dimensional features. Finally, we adopt 3 fully connected layers to combine the low-level features into higher-level ones and output the predicted redshift.
\item {\bf GaZNet$-$C4}. This is a simple ANN structure with 2 blocks made of 4 fully connected layers, separated by a flatten layer. The input is an optical 4-band ($u,g,r,i$) catalog of magnitude and color measurements. Since we use the information from the same bands, comparing GaZNet$-$I4 and GaZNet$-$C4 allows us to quantify the impact of the imaging and photometry on the redshift estimates.
\item {\bf GaZNet$-$C9}. It has the same structure as GaZNet$-$C4, but it is input with the 4-band optical catalogs from KiDS plus the 5-band catalogs from the VISTA Kilo-degree Infrared Galaxy survey (VIKING, \citealt{Edge+14_VIKING-DR1}, see \S\ref{sec:train_test_data} for details).
%The structure of the ML model is totally the same as GaZNet$-$C4. 
%For the inputs, apart from the 4-band optical catalogs from KiDS, we add another 5-band catalogs collected from the VISTA Kilo-degree Infrared Galaxy survey (VIKING, \citealt{Edge+14_VIKING-DR1}, see Sec \ref{sec:train_test_data} for detail). 
Using a broader wavelength baseline, GaZNet$-$C9 %will make use of all of the available catalogs for KiDS galaxies to achieve high 
will allow us to estimate the impact of the multi-band coverage on the ANN redshift predictions.
%accurate \hai{photo-$z$}.
%\item {\bf GaZNet$-$C}: This is a simple MPL structure with 2 blocks and in each block there are 4 fully connected layers. Between the two blocs, a flatten layer is added. Depending on how many band are input, we train two models for GaZNet$-$C. The first one, labeled as GaZNet$-$C4, is input with 4-band ($u,g,r,i$) catalog of magnitudes and the driven colors, while the second one, labled as GaZNet$-$C9, is input with 9-band catalogs. GaZNet$-$C4 allow us to know whether the images or the catalogs can provide more information for photz determination. GaZNet$-$C9 will make use of all of the catalog information in KiDS to achieve a high accurate photz.
\item {\bf GaZNet-1}. This is the reference network we have developed: the input is the combination of the $r$-band images and the multi-band photometry catalogs. The GaZNet-1 has been designed to have a two-path structure.
The first path comprises 4 blocks as GaZNet$-$I4, while the second is made of 8 fully connected layers as GaZNet$-$C4 and C9. After a flatten layer, the features from each path are concentrated together. Finally, 5 fully connected layers are added to combine the features from images and catalogs to generate %output the \hai{photo-$z$}
the final redshift predictions.
\end{itemize}
%We note that only GaZNet-1 is the expected tool in this work and will be finally applied to KiDS galaxies in the future for studying the evolution, while other 3 ML tools are designed only for comparison and tests.
Of the four tools illustrated above, the first three are mainly designed to test the impact of the different inputs on the final redshift estimates. Being constructed with the same structure assembled in the final GaZNet-1, i.e. the one to be used for science, they guarantee the homogeneity of the treatment of the input data (see Fig. \ref{fig:CNN_model}). 
%which are handled in the same way they are used from the morphot-$z$ regressor represented by the GaZNet.

In this first series of GaZNets, we do not consider the magnitude ratios between different bands as inputs, although there are experiments suggesting that they can improve the precision (see e.g., \citealt{2018A&A...616A..97D}, \citealt{2019A&A...624A..13N}). We plan to implement this in future analyses, because here we are interested in checking the advantages of the combination of images and photometry compared to previous analyses made on the same data (see \S\ref{sec:test_external}).
Also, in this analysis, we focus on redshift point estimates. In the future, we plan to expand the capabilities of the GaZNets to estimate the probability density function $p(z)$ for each galaxy.
%, instead of the single point output of photo-z. 
This can be achieved by Mixture Density Networks (e.g., \citealt{ Rhea2021RNAAS...5..276R}, \citealt{Wang2022arXiv220700185W}) or Bayesian Networks (\citealt{Gal2015arXiv150602158G}; \citealt{Kendall2017arXiv170304977K}), and some works with these two networks have been done (e.g.,\citealt{D'Isanto2018A&A...609A.111D}; \citealt{ Ramachandra2021arXiv211112118R}; \citealt{Podsztavek2022ascl.soft04004P})
%examples of their applications on the photo-zs are SYTH-Z (\citealt{ Ramachandra2021arXiv211112118R}) and SZNet (\citealt{Podsztavek2022ascl.soft04004P}). 
We will also evaluate the performance of the $p(z)$ using cumulative distribution function (CDF) and CDF-based metrics, such as the Kolmogorov–Smirnov (KS) statistic, the Cramer-von Mises statistic and the Anderson–Darling (AD) statistic (see details in \citealt{Schmidt2020MNRAS.499.1587S}).

\subsection{Training and testing data}
\label{sec:train_test_data}

The dataset used in this work is collected from KiDS and VIKING, two twin surveys covering the same $1350$ deg$^2$ sky area, in optical and NIR bands, respectively. %\hai{ KiDS and VIKING.}
%the Kilo Degree survey (KiDS) and the VISTA Kilo-degree Infrared Galaxy (VIKING, \citealt{Edge+14_VIKING-DR1}) survey.
KiDS observations are carried out with the VST/Omegacam telescope (\citealt{Capaccioli2011Msngr.146....2C}; \citealt{Kuijken2011Msngr.146....8K})
%is carried by the European Southern Observatory (ESO), aiming to image 
%$~1350$ $Deg^2$ sky 
in 4 optical filters ($u, g, r, i$), with a spatial resolution of 0.2$''$/pixel. The $r$-band images are observed with the best seeing (average FWHM$\sim0.7''$), and its mean limiting AB magnitude ($5\sigma$ in a $2''$ aperture)  is $25.02\pm0.13$. The seeing of the other 3 bands ($u, g$ and $i$) is slightly worse than that of the $r$-band i.e. FWHMs $<1.1''$, and the mean limiting AB magnitudes are also fainter, i.e. $24.23\pm0.12$, $25.12\pm0.14$, $23.68\pm0.27$ for $u, g$ and $i$, respectively (\citealt{Kuijken+19_KiDS-DR4}). 
%With these deep and high quality images, people can got not only the accurate magnitudes for galaxies, but also the parameters on light distributions (e.g. \citealt{Roy+18}). 

VIKING is carried out with the VISTA/VIRCAM (\citealt{Sutherland2015A&A...575A..25S_VISTA}) and aims at
%is a complemented survey for KiDS, aiming taking 
complementing KiDS observations with five NIR bands ($Z, Y, J, H$ and $K$s). The  median value of the seeing in the images is $\sim 0.9''$ (\citealt{Sutherland2015A&A...575A..25S_VISTA}), and the AB magnitude depths are 23.1, 22.3, 22.1, 21.5 and 21.2 in the five bands (\citealt{Edge2013Msngr_VIKING}), respectively.
%for the same area of KiDS.

In particular, the galaxy sample used in this work is made of 
%We collect 
148\,521 objects 
%galaxies in KiDS+VIKING, and the 
for which spectroscopic redshifts (spec-$z$s, hereafter) 
%of these galaxies have been obtained 
are available from different surveys, such as the Galaxy And Mass Assembly survey (GAMA, \citealt{Driver2011MNRAS_GAMA}) data release 2 and 3, the zCOSMOS (\citealt{Lilly2007ApJS_zcosmos}), the  Chandra Deep Field South (CDFS, \citealt{Szokoly2004ApJS_CDFS}), and the DEEP2 Galaxy Redshift Survey (\citealt{Newman2013ApJS_DEEP2}). 
The spec-$z$ range of the galaxies
%The redshift range of the spectroscopic sample
covers quite a large baseline, between $\sim 0-7$, although the distribution is far to be uniform.
Indeed, as shown in Fig. \ref{fig:para_distri}, the number of galaxies at higher redshift ($z\gtrsim0.8$) is much smaller than the one at lower redshift. In the same figure,
we can see a peak of distribution at spec-$z$ $<0.6$. It comes from the GAMA survey,
%we can see the peak of redshifts at spec-$z$ $<0.6$ coming from the GAMA survey,
which is
%also
the most complete 
%of the 
spectroscopic surveys adopted, with $\sim95.5\%$ completeness for $r$-band magnitude MAG$\_$AUTO$<19.8$ (\citealt{Baldry2018MNRAS_GAMADR3}).
Similarly, we see a second peak at spec-$z$ $\sim2.5$, due to the quite deep observations from zCOSMOS.
Overall, this sample is dominated by bright and low redshift galaxies ($0.04<z<0.8$), however, between $0.8<z<3$, it still contains $\sim6500$ galaxies with a quite uniform redshift distribution that can be used as training sample to extend the predictions to higher redshift.
%Overall, the redshift coverage in the range of $0.04<z<3$ (see Fig. \ref{fig:para_distri}) shows a relatively smooth distribution.
Due to the unbalanced redshift coverage, we expect the accuracy of the predictions to have a strong variation with redshift. However, we will check if the final estimates meet the accuracy and precision requirements
for cosmological and galaxy formation studies (see e.g., \citealt{LSST2009arXiv0912.0201L}).
%study in the range of $0.04<z<3$ (see Fig. \ref{fig:para_distri}), which result in 
After this redshift cut, the final sample is made of 134\,148 galaxies. The distributions of the r-band Kron-like magnitude, MAG$\_$AUTO, obtained by SExtractor \citep{Bertin_Arnouts96_SEx} for these galaxies, and their signal-to-noise ratio (SNR, defined as the inverse value of the error of MAG$\_$AUTO, are also reported in Fig. \ref{fig:para_distri}. 
%These two latter quantities 
%where they both show a characteristic log-normal distribution.
Finally, the 134\,148 galaxies are 
%further classified 
divided into three datasets, 100\,000 for training, 14\,148 for validation, and 20,000 for testing and error statistical analysis.

The $u, g, r, i$ band images, with size of $8\times8''$, are cutout from KiDS DR4 (\citealt{Kuijken+19_KiDS-DR4}). The corresponding catalogs, made up of 9 Gaussian Aperture and point spread function (GAaP) magnitudes ($u, g, r, i, Z, Y, J, H, K$s) and 8 derived colors (e.g., $u-g$, $g-r$, $r-i$ etc.), are directly selected from the KiDS public catalog\footnote{https://kids.strw.leidenuniv.nl/DR4/access.php}. The GAaP magnitudes have been measured on Gaussian-weighted apertures, modified per-source and per-image, therefore providing seeing-independent flux estimates across different observations/bands, reducing the bias of colors (see detail in \citealt{Kuijken2015MNRAS,Kuijken+19_KiDS-DR4}). The extinction was also considered in the measurement of the GAaP magnitudes.

\subsection{External photo-$z$ catalog by MLPQNA}
\label{sec:external_MLPQNA}
To test the performances of GaZNet-1 against other ML based photo-$z$ methods, we have collected an external photo-$z$ catalog obtained from MLPQNA for the same KiDS galaxies we have used as testing sample. This allows us to perform a quantitative comparison of diagnostics like accuracy, scatter, and fractions of outliers.

MLPQNA is an effective computing implementation of neural networks adopted for the first time to solve regression problems in the astrophysical context. A test on PHAT1 dataset (\citealt{Hildebrandt2010A&A}) indicated that MLPQNA, with smaller bias and fewer outliers, performs better than most of the traditional standard SED fitting methods. This code has been used in some current sky surveys, e.g., KiDS (\citealt{Cavuoti+15_KIDS_I}) and Sloan Digital Sky Survey (SDSS; \citealt{Brescia2014A&A}). For our comparison, we adopt the MLPQNA photo-$z$ catalog from \cite{Amaro2021+photz}, where they have used the same data presented in \S\ref{sec:train_test_data} to train and test their networks.

\section{GaZNet training and testing}
In \S\ref{sec:arch} we have described the different GaZNets and anticipated that they accept either images or catalogs of galaxies as inputs, except the GaZNet-1, fed with both images and catalogs. In particular, 
for the first test of morphoto-$z$ predictions
%for the first morphoto-$z$ prediction test
made with this latter, we choose only the $r$-band images, i.e. the ones with best quality
%best quality $r$-band images 
from KiDS, to combine with the 9-band photometry catalog. As we will demonstrate in \S\ref{sec:discussion}, the multi-band imaging does not add sensitive improvements in the results for the higher computation time prize one has to pay.  

In this section, we illustrate the procedures to train the networks and test their predicted photo-$z$s
%predicted redshifts 
against the ground truth provided by the spec-$z$s of the test sample introduced in \S\ref{sec:train_test_data}.

\subsection{Training the networks}

\begin{figure*}
\centerline{\includegraphics[width=18cm]{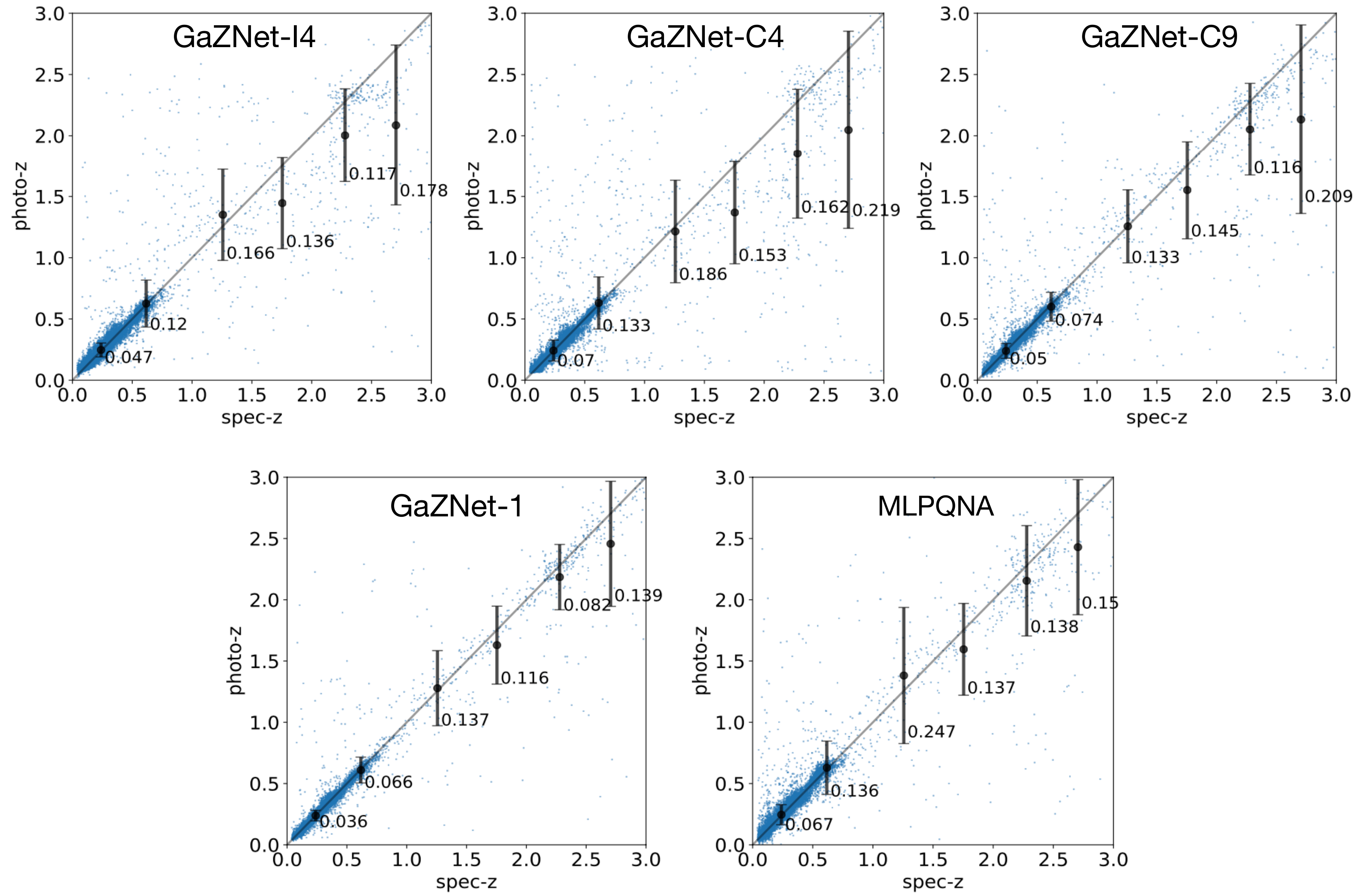}}
\caption{Comparison between the spectroscopic redshifts and the predicted photometric redshifts for different models.  From top left to bottom right are the results from GaZNet-I, GaZNet-C, GaZNet and MLPQNA, respectively. Error bars represent the mean absolute errors (MAE), while the quoted numbers are the mean $|\delta z|$, in each bin.}
\label{fig:comparison}
%\vspace{1cm}
\end{figure*}

%and the outputs are the photometry redshift. 
We train the networks by minimizing the ``Huber" loss (see, \citealt{Huber10.1214/aoms/1177703732}; \citealt{Friedman99+huberloss}) function with an ``Adam" optimizer (\citealt{Kingma2014+Adam}). The ``Huber'' loss is defined as
\begin{equation}
	L_{\delta}(a) =
	\begin{cases}
	    \dfrac{1}{2}(a)^2,  \ \ \ |a|\leq\delta\\
	    \delta\cdot(|a|-\dfrac{1}{2} \delta),  \ \ \ {\rm otherwise}.
    \end{cases}
\end{equation}
In which $a=y_{\rm true}-y_{\rm pred}$. $y_{\rm true}$ is the spec-$z$ and $y_{\rm pred}$ is the predicted photo-$z$. $\delta$ is a parameter that can be pre-set. Given a $\delta$ (fixed to be $0.001$ in this work), the loss will be a square error when the 
deviation of the prediction, $|a|$,
%prediction deviation $|a|$ 
is smaller than $\delta$; otherwise, the loss is reduced to a linear function. Compared to the commonly used Mean Square Error (MSE) or Mean Absolute Error (MAE) loss function defined as
\begin{equation}
\begin{aligned}
	{\rm MAE}=\frac{1}{n}\sum |z_{\rm pred}-z_{\rm spec}| \ \ 
	\\
	{\rm MSE}=\frac{1}{n}\sum (z_{\rm pred}-z_{\rm spec})^2.
\end{aligned}
\end{equation}
``Huber" loss is proved to be more accurate 
in
%for 
such regression tasks (see detail discussion in \citealt{Li2021_GaLNet}).

To guarantee a faster-reducing speed of the loss function, for each ML model, we set a larger learning rate of 0.001 at the beginning and train the networks for 30 epochs. In each epoch, the networks are trained on the training data and validated on the validation data to verify if further adjustments are needed to improve the overall accuracy. After the first training round, we reduce the learning rate to 0.0001, and load the pre-trained model with a ``callback'' operation. Then we train the networks for further 30 epochs. Changing the learning rate to a smaller value of 0.0001
%was found to 
can help 
%to let 
the network converge to the global minimum, hence finding the best-trained model.

For the networks input with images, we also apply some data augmentations, including random shift, flip, and rotation (only $90^{\circ}, 180^{\circ}$ and  $270^{\circ}$). We do not adopt any augmentation that needs interpolation algorithms\footnote{Note that, even if a generic rotation does imply some interpolation due to pixel re-sampling, the adoption of $\pi/2$ multiples does not, because it preserves the overall geometry of the cutout, except the orientation.}, like crop, zoom, color-changing, and adding noise, since these operations would change the flux in the image pixels, affecting the magnitudes and colors of the galaxies.

Computing-time wise, with the NVIDIA RTX 2070 graphics processing unit (GPU), GaZNet-C4 and GaZNet-C9 need $28$ minutes to complete the training and validation process, while GaZnet-1 takes about 134 minutes (including $\sim2.5$ minutes for data reading) because of the time needed to process the r-band image data along with the magnitudes and colors. Compared with GaZNet-1, GaZNet-I4 does not significantly increase the time during the training and validation process. However, to deal with the 4 channel images, GaZNet need much larger memory of GPU, and time of data reading increases significantly.
%and requirement on the GPU memory also becomes much higher. 
Finally it needs $\sim 4$ times longer on data reading. We also estimated that to make predictions on 20\,000 galaxies (see \S \ref{sec:testing}), in terms of the computing time for predictions, this is much shorter. GaZNet-C9 needs only $\sim 6$ seconds for the whole galaxies, while GaZNet-1 needs $\sim 55$ seconds, including $\sim 30$ seconds for data reading and $\sim 25$ seconds for prediction.

\subsection{Testing the performance}
\label{sec:testing}

%\begin{figure*}
%\centerline{\includegraphics[width=18cm]{pdfs.pdf}}
%\caption{Comparison of the PDFs GaZNet, GaZNet-I, GaZNetC and MLPQNA. %respectively. The spectroscopic redshift distribution is show with blue %bars.}
%\label{fig:pdfs}
%\vspace{1cm}
%\end{figure*}

After the training phase, we use the 20\,000 testing galaxies
%of the test sample 
to estimate the precision 
%the ML tools 
and the statistical errors of the redshift predictions from different GaZNets. 
%Here 
\subsubsection{Statistical parameters}
\label{sec:Statistical_parameters}
We define a series of statistical parameters to describe the overall performances:  1) the fraction of catastrophic outliers, 2) the mean bias, and  3) the normalized median absolute deviation (NMAD).

%\rui{The catastrophic outliers are defined as the galaxies with relative errors larger than 15\% according to the following formula::}
The fraction of the catastrophic outliers is defined as the fraction of galaxies with bias larger than 15\% according to the following formula:
\begin{equation}
\label{fuc:delta_z}
	|\delta z|=\frac{|z_{\rm pred}-z_{\rm spec}|}{1+z_{\rm spec}}>15\%.
\end{equation}
where $z_{\rm spec}$ are the spec-$z$s of the test galaxies and $z_{\rm pred}$ are the predicted redshifts by the ML tools. This definition is usually adopted for outliers in photo-$z$ estimates (see details in e.g., \citealt{Cavuoti2012A&A_MLPQNA}, \citealt{Amaro2021+photz}) and gives a measure of the fallibility of the method. In addition, the mean bias in this work is labeled as $\mu_{\delta z}$.

The normalized median absolute deviation (NMAD), between the predicted photo-$z$s and the true spec-$z$s,
%spectroscopic redshift 
are defined as
\begin{equation}
\begin{aligned} 
	{\rm NMAD}=1.4826\times {\rm median} (|\delta z - {\rm median }(\delta z)|),
\end{aligned}
\end{equation}
where $\delta z$ comes from Eq. \ref{fuc:delta_z}. NMAD allows us to quantify %the amount of 
the scatter of the overall predictions in comparison to the ground truth, hence it is a measurement of the precision %precision
of the redshift estimates from the ML tools.

%Finally, we use a linear fit to interpolate the $\delta z$ variation as a function of $z$,
%\begin{equation}
% \delta z=A (1+z),
%\end{equation}
%where $A$ is the slope to be fitted from the data.
%This relationship will give an overall  precision changing with redshift for the different tools.

\subsubsection{Predictions vs. ground truth}
\label{sec:perform}
The testing results on 20\,000 galaxies for the four GaZNets are shown in Fig. \ref{fig:comparison}, where on the x-axis we plot the spec-$z$s as ground truth, and on the y-axis we plot the predicted redshifts. As a comparison, in the same figure, we also show the photo-$z$s estimated by the MLPQNA. We divided the galaxies into 6 redshift bins, and computed the mean absolute errors, shown as error bars, and the mean $|\delta z|$ defined in Eq. \ref{fuc:delta_z}, reported as text. We use equally spaced bins to check the effect of the sampling as a function of the redshift.

From Fig. \ref{fig:comparison}, a major feature one can see, at the first glance,
%general comment is about 
is the odd coverage of the spec-$z$ at high redshift ($z\gtrsim0.8$),
%over the whole redshift range, 
which we have also discussed in \S\ref{sec:train_test_data}. This is a potential issue for all methods, as a poor training set can introduce a large scatter in the predictions. Indeed, 
%This is a feature that we can see for all methods 
in Fig. \ref{fig:comparison} the $\delta_z$ tends to have an increasingly larger scatter at larger redshifts. This means that at $z\gtrsim0.8$ the absolute scatters are dominated by the size of the training sample rather than the true intrinsic uncertainties of the methods. Unfortunately, this is a problem we cannot overcome with the current data 
%we have on hand, 
and we need 
%to postpone to forthcoming analyses, when 
to wait for larger spec-$z$ data samples
%spec-$z$ data samples 
to improve the precision at higher redshifts.
%will be available. 
However, given the current training set, we can still evaluate the relative performances of different methods and their ability to make accurate predictions even in the small training set regimes. 

Given this necessary preamble, from Fig. \ref{fig:comparison} we see that unbiased photo-$z$ can be obtained by GaZNet-I4, with only 4-band images as input,
%in which only $4-$band optical images are input, 
although it seems that this starts to deviate from the one-to-one relation at $z\gtrsim1.5$. However, at these redshifts,
%we see that 
a general trend of underestimating the ground truth is also shown by GaZNet-C4 and GaZNet-C9, although the latter uses the full photometry from the KiDS+VIKING dataset. Interestingly, looking at the scatter, GaZNet-I4 seems to perform better than GaZNet-C4 at all redshift bins and almost comparably to C9 in most cases.

%{\it To our knowledge, this is the first time that it is demonstrated that morpho-$z$s from multi-band imaging are similar, if not potentially superior to photo-$z$s from photometry in the same bands.} 
%The result of GaZNet-I4 
The results from GaZNet-I4 
demonstrate that morpho-$z$s from multi-band images are similar, if not potentially superior, to photo-$z$s from photometry in the same bands.
This high-performance of morpho-$z$s is also confirmed by the noticeably smaller outlier fraction (1.5\% for GaZNet-I4 and 2.2\% for GaZNet-C4 in general). Even more interestingly, looking from the perspective of future surveys relying on a narrower wavelength baseline, such as the space missions Euclid and CSST, our results show that morpho-$z$s are not far from optical+NIR large photometric baselines in terms of accuracy, scatter, and the fraction of outliers. This is particularly true for $z<1$, where, as seen in Fig. \ref{fig:comparison}, the GaZNet-I4 shows less outliers than GaZNet-C9, while this latter shows a rather lower fraction of outliers at higher redshifts.

Moving to GaZNet-C9, the results show the impact of the broader wavelength baseline including the five NIR bands. Generally, photo-$z$ determined by GaZNet-C9 show improved indicators in comparison to GaZNet-I4 and GaZNet-C4. From Fig. \ref{fig:comparison} we can see these coming from a better linear correlation, especially at $z>1.5$, and smaller absolute errors.
However, looking at the results in more detail,
%from Fig. \ref{fig:comparison} one can see indeed that the larger $R^2$ comes from a better linear correlation, especially at $z>1.5$. However, in this redshift range, 
at $z>1.5$ the presence of a rather large fraction of outliers causes the median values to diverge from the one-to-one relationship in a way similar to GaZNet-I4 and GaZNet-C4. It is hard to assess whether this is caused by the poor training sample, or it is an intrinsic shortcoming of the ML tool. In either cases, it is important to check whether using the information from images can improve this result.

Compared to GaZNet-C9, GaZNet-1 has overall better performances, with a tighter one-to-one relation, and smaller errors (by $\sim10-35$\%) in all redshift bins. This is shown in the bottom-left panel of Fig \ref{fig:comparison}. This result leads us to two main conclusions: 1) images (even a single high-quality band, see \S\ref{sec:discussion} for the test on multi-band imaging) provide crucial information to solve intrinsic issues related to the photometry only 
%tools
and improve all performances of the redshift predictions, in terms of accuracy, scatter, and outlier fraction. We will discuss the reason for this in \S\ref{sec:discussion}; 2) due to the poor redshift coverage of the training sample at $z>1$, the results we have obtained possibly represent a lower limit on the potential performances of the tool. No matter what, the GaZNet-1 reaches an excellent overall precision of $\delta_z=0.038(1+z)$ up to $z=3$, with an overall outlier fraction of $0.74$\%. 

\begin{figure*}
\centerline{\includegraphics[width=18cm]{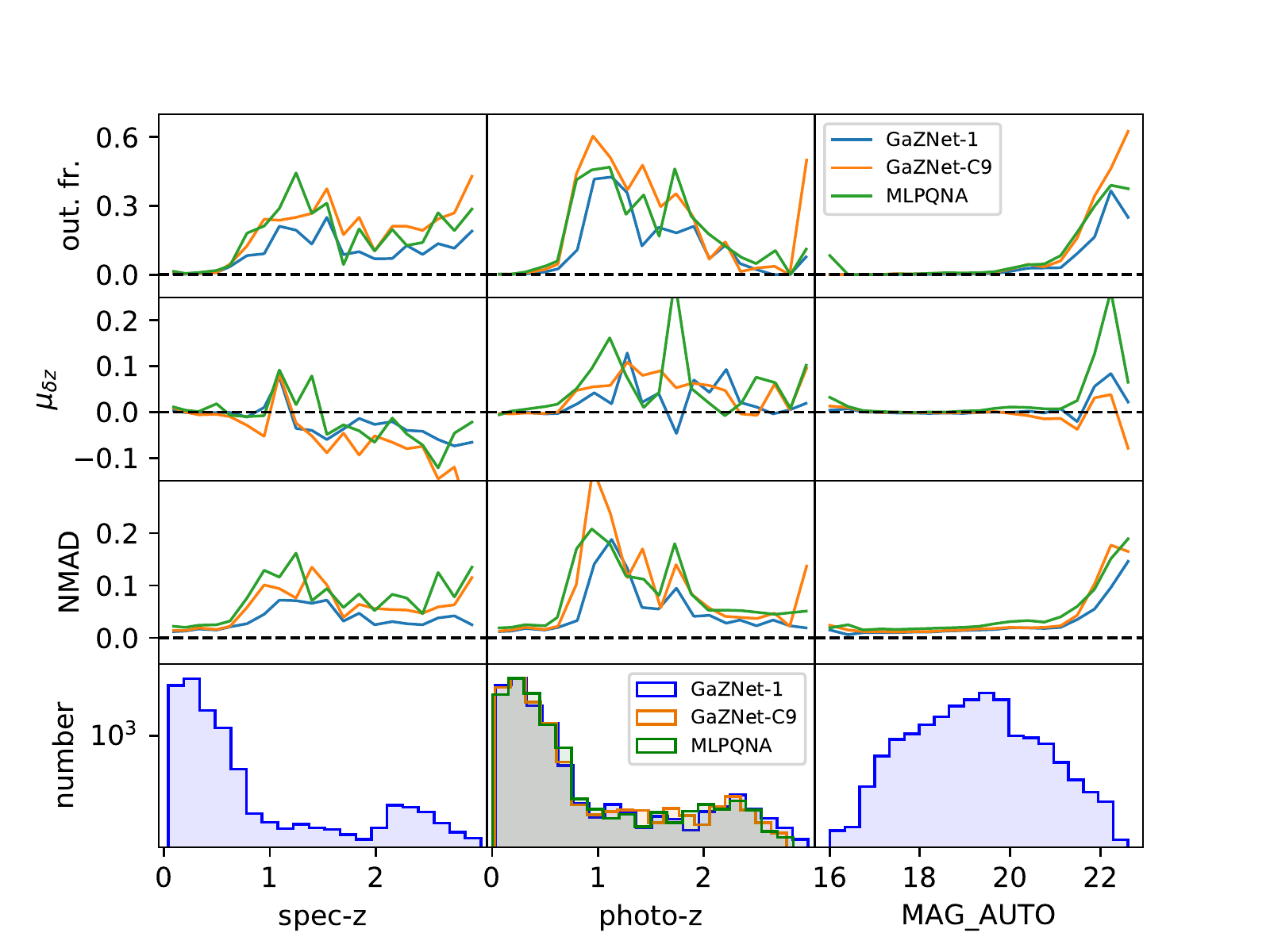}}
\caption{Outlier fraction (out. fr.),  mean bias ($\mu_{\delta z}$), and Scatter (NMAD) as functions of spec-z, photo-z, and magnitudes in 20 bins. In each panel, blue line is for GaZNet-1, orange is for GaZNet-C9 and green is for MLPQNA. In the last row we also present the number distribution in the corresponding parameter space.}
\label{fig:statistics_of_bias}
%\vspace{1cm}
\end{figure*}

\subsubsection{Test vs. external catalogs}
\label{sec:test_external}
%These performances are not even closely equalized by the external catalogs
We can finally compare the performance of the four GaZNets versus the external catalogs.
The MLPQNA is rather similar to the ANN
%MLP 
method used for the GaZNets-C9, as it makes use of a similar algorithm and the same catalogs from KiDS DR4. From Fig. \ref{fig:comparison} we see that MLPQNA performs similar to GaZNet-I4 and better than GaZNet-C4. This is not surprising as the MLPQNA uses a larger wavelength baseline. This is particularly visible at
higher redshift, where the predictions from MLPQNA are tighter distributed around the one-to-one relationship with the ground truth than GaZNet-I4 and GaZNet-C4. 

\subsubsection{Performance in space of redshift and magnitude }
\label{sec:parameter_space}
In the last subsections, we have shown that GaZNet-1 shows better performances than others tools, in terms of accuracy and precision. However, in Fig. \ref{fig:comparison}, we have also seen a variation of these performances as a function of the redshift. Here, we want to more detailedly quantify 
%in more details 
this effect as well as the 
%and also evaluate 
the dependence on the magnitudes of the same performances. The reason for this diagnostics is to assess the impact of selection effects on the overall performances (see e.g. \citealt{Busch2020A&A...642A.200V}). E.g., in \S\ref{sec:perform}, we have anticipated that the redshift sampling by the training sample can be one source of degradation of the performances at $z>1$.

\begin{table}
\footnotesize
\begin{center}
\caption{\label{tb:parametersl} Statistical properties of the predictions.}
\begin{tabular}{c c c c}
\hline \hline
CNN model & Out. fr. &$\mu_{\delta z}$& NMAD \\
 \hline
\multicolumn{4}{c}{Low redshit galaxies ($z\le0.8$)}\\
 \hline
GaZNet-C9& 0.007 & 0.0& 0.016 \\
GaZNet-1 & 0.004& 0.0& 0.014 \\
MLPQNA  & 0.01& 0.006& 0.022 \\
\hline
\multicolumn{4}{c}{high redshift galaxies  ($0.8<z<3$)} \\
\hline
GaZNet-C9& 0.234& -0.067& 0.073\\
GaZNet-1& 0.127& -0.028& 0.041\\
MLPQNA & 0.216& -0.022& 0.087\\
\hline \hline
\end{tabular}
\end{center}
\begin{flushleft}
\textsc{Note.} ---Outlier fraction, mean bias and NMAD (see \S \ref{sec:Statistical_parameters}) from different tools on lower ($z<0.8$) and higher (z>0.8) redshift galaxies. 
%From left to right, we show the fraction of outliers, the mean bias and the NMAD as defined in \S \ref{sec:Statistical_parameters}.}
\vspace{0.3cm}
\end{flushleft}
\end{table}

%in the space of 
%as a function of redshift and magnitude. 
In Fig. \ref{fig:statistics_of_bias} we plot the outlier fraction (out. fr.), 
mean bias ($\mu_{\delta z}$) and scatter (NMAD) as functions of spec-z, photo-z, and $r$-band magnitudes. 
%These diagnostics, as shown in previous analyses (e.g. \citealt{Busch2020A&A...642A.200V}), allow us to mitigate the impact of the selection effects in the performance assessment. 
%this can make the analysis of the results less susceptible to selection effects. 
As comparison we plot the same relations for GaZNet-C9 and MLPQNA, the other two tools showing comparable performances to the GaZNet-1.
%, whose performance are second and third only to the reference tool as shown in previous sections. 
The bottom row of the same Fig. \ref{fig:statistics_of_bias}, finally shows the distribution of the training sample
%, randomly extracted from the original training sample, 
in the same parameter space.
%are also shown in the last row of the same figure. 
%Here we note that the testing galaxies are randomly selected from the whole sample, therefore, the distribution of the testing galaxies and the training galaxies are the same.

From this figure, the overall impression is that %general impression is that 
%the reference tool 
GaZNet-1 performs generally better than the other two tools in most, if not all, redshift and magnitude bins, with lower outlier fraction, smaller mean bias and scatter.
%almost all of the bins.
We also see a clear correlation of the performances of all tools, included GaZNet-1, 
%as well as the other two tools, are mainly affected by two factors: the galaxy 
with the size and magnitude of the training sample in different redshift bins and the magnitudes of the training galaxies. All of the 3 tools performs quite well in the range of $z\lesssim0.8$, where the training sample is about one order of magnitude larger, resulting on a more accurate training.
To quantify the overall performance in this redshift range,
%for this area, 
in Tab \ref{tb:parametersl} we report the global statistical parameters for these galaxies. The three tools all can achieve quite small outlier fractions ($\lesssim0.01$), mean bias (close to 0) and scatters ($\lesssim 0.022$). 
%In addition, the estimated NMADs ($\lesssim 0.022$), are well below the requirements for weak gravitational lensing studies in next generation ground-base surveys (e.g., NMAD=0.05 in VR/LSST, \citealt{LSST2009arXiv0912.0201L}).
Compared to the other two tools, GaZNet-1 shows the best performance. In particular, its outlier fraction is $43\%$ smaller than GaZNet-C9 and $60\%$ smaller than MLPQNA.
%we have enough galaxies to train the ML tools.

At $z\gtrsim 0.8$, in Fig. \ref{fig:statistics_of_bias} we see that the number of galaxies decreases rapidly, which produces a degradation of the performances of all tools.
%making the measured performances become worse. 
Interestingly, looking at the central panels,
%This worse performance continues until $z\sim 2$, 
after $z\sim 1.5$, where the COSMOS spec-$z$ sample 
%provides a rather numerous amount of spec-$z$, 
is concentrated, the performances, especially in terms of scatter (NMAD), show a significant improvement up to $z\gtrsim 2.6$, where the spec-$z$ of the training sample quickly drop in number again. This is also quantified In Tab \ref{tb:parametersl} 
%we also present the 
by the global statistical parameters for these higher redshift galaxies.
%at $0.8<z<3$. 
Compared to GaZNet-C9, all the indicators from GaZNet-1 are significantly improved. The fraction of outliers, the mean bias and the scatter are decreased by $46\%$, $58\%$, and $44\%$, respectively. On the other hand, MLPQNA remains the tool performing worse.
%Among the three tools, only the scatter from GaZNet-1 (NMAD=0.041) is below the requirements for weak gravitational lensing in LSST.

A similar behavior of the performances is seen also as a function of the photo-$z$, as these latter closely follow the spec-$z$ (see Fig. \ref{fig:comparison}).

Going to the magnitude space, we find that all indicators shows very small values at MAG$\_$AUTO $\lesssim 21$, which means that the redshift estimates for brighter galaxies are highly reliable. 
After $r$-band MAG$\_$AUTO$\sim 21$, 
%the values of 
all indicators degrade, 
%toward higher values, 
showing a worsening of the accuracy, precision and outlier fraction. Among the three tools, the GaZNet-1 is the one with better performances, though.
In particular, after MAG$\_$AUTO$\sim 22$ there is a peak at $30\%$, which can be either driven by the poorer SNR of the systems, but more likely by the smaller statistics. 
In general, though, the GaZNet-1
%Looking specially on GaZNet-1, 
has still a quite small outlier fraction ($\sim1.0$\%).
%, except two cases: $0.8\lesssim z \lesssim2$ and MAG$\_$AUTO $\gtrsim22$. For the former, we can safely attribute the worse performance to the lack of the training galaxies. For the later, we still don't know if it is because the galaxies are too faint or because there are too few samples. 

Overall, it is clear that collecting more galaxies covering higher redshift ($z\gtrsim 0.8$) and fainter magnitudes (MAG$\_$AUTO $\gtrsim22$) will be essential for improving the performance of these ML tools, and the results collected here represents just a lower limit of the real performances that these tools, especially GaZNet-1, can achieve.  
However, even with the current training set, GaZNet-1 can provide even for high redshift galaxies results that satisfy the requirements for weak gravitational lensing studies in next generation ground-base surveys (e.g., NMAD=0.05 in VR/LSST, \citealt{LSST2009arXiv0912.0201L}), although this has been currently tested only on a relatively bright sample with AB magnitude MAG$\_$AUTO$\leq22$ (see \ref{fig:statistics_of_bias}). For the low-redshift samples the GaZNet-1 is already well within the requirements for the same surveys and it is virtually science ready.
In the future, we will look for more higher redshift galaxies from different spectroscopic surveys to build a less biased training sample and improve the performances at $z>0.8$.

\section{Discussion}
In the previous section, we have compared the performances of the different GaZNets based on different architectures, including or not deep learning. We have also compared the GaZNets against external catalogs of photo-$z$ based on traditional machine learning algorithms. The reason for us to develop different tools with an increasing degree of complexity, is to understand the impact of the different features in the final predictions. The main conclusion of this comparison is that GaZNet-1, using the combination
%combined information 
of 9-band photometry and $r$-band imaging, clearly over-perform all the other tools, either developed by us or taken from literature, based on photometry only and no use of deep learning. We have also seen how deep learning only, applied to only 4-band optical images, can produce morpho-$z$ that are more accurate than the photo-$z$ from the corresponding photometry and equalize the performance of the 9-band photometry, except for redshift larger than $z>0.8$. Overall, we have discussed that part of the over-performance of deep learning is concentrated on the outlier fraction. In this section, we investigate the reasons for these results and discuss the impact of some choices we have made in the set-up of the GaZNets presented in this first paper.   
\label{sec:discussion}
\subsection{The outliers}

\begin{figure*}
\centerline{\includegraphics[width=16cm]{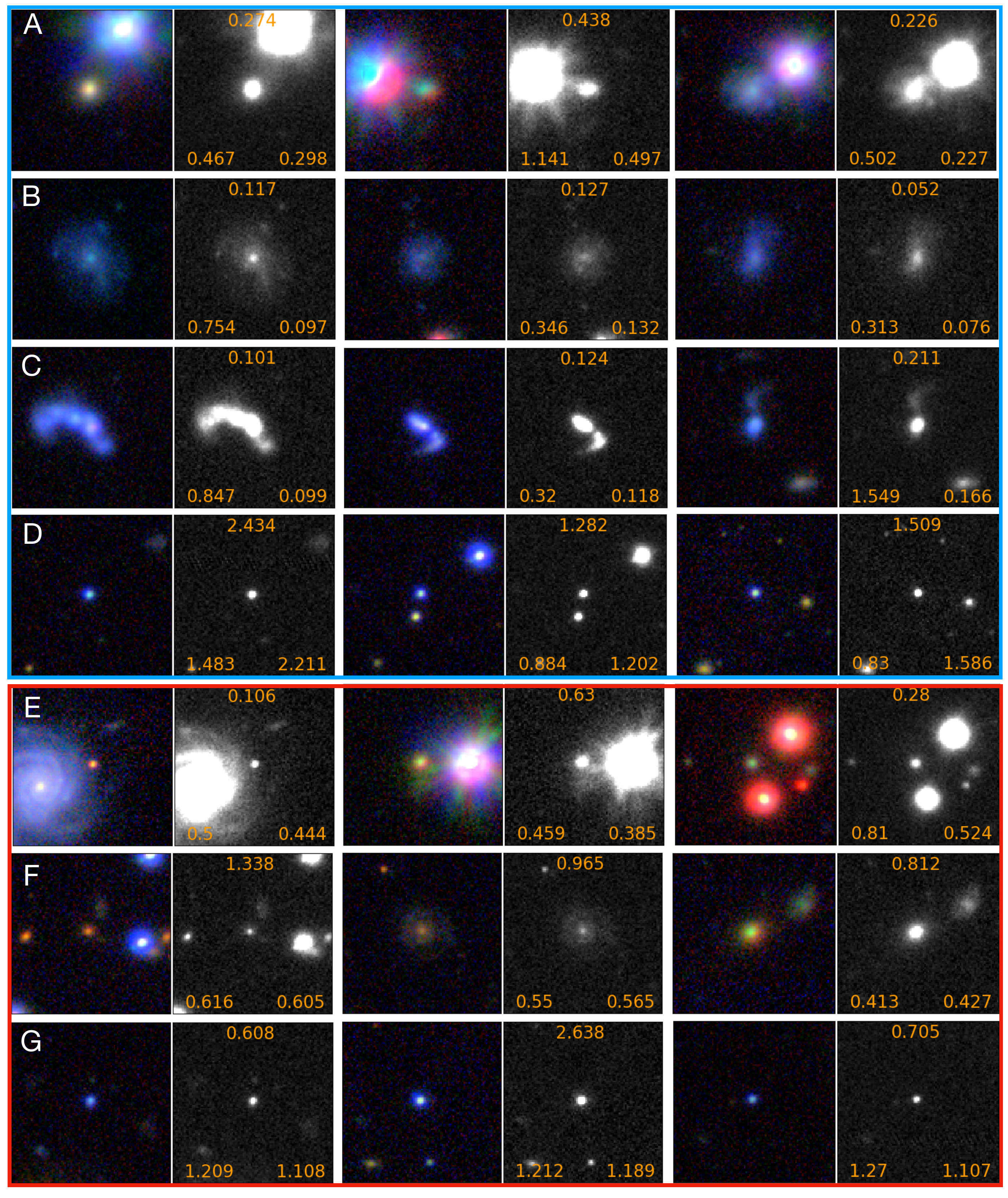}}
\caption{The $g, r, i$ color-composited color
images ($20''\times20''$) and the corresponding $r$-band images for some representative outliers. Rows A, B, C and D (blue framed) show the outliers in GaZNet--C9, which are no longer outliers in GaZNet-1 predictions. Rows E, F and G (red framed) show the objects that remain outliers both for GaZNet-1 and GaZNet-C9. In the $r-$band images we report the spec-$z$ on the top, the GaZNet-C9 photo-$z$ on the bottom left and the GaZNet-1 morphoto-$z$ on the bottom right.}
\label{fig:outliers}
%\vspace{1cm}
\end{figure*}

From Table \ref{tb:parametersl}, it is evident that the major advantage from deep learning, applied to high-quality imaging, resides in the low outlier fraction. For GaZNet-1, this is smaller than the one of GaZNet-C9 by $\sim43\%$ for low redshift galaxies ($z<0.8$) and $46\%$ for higher redshift galaxies ($z>0.8$). Understanding the reasons for these results is important to %help us to 
figure out the source of systematics and plan next developments for more accurate morphoto-$z$ estimates.

To investigate the genesis of these outliers, we start by checking the galaxies for which the
%the kind of galaxies the 
ML tools fail to obtain accurate photo-$z$s.
In Fig. \ref{fig:outliers}, we show the optical $gri$ color-composed and the $r$-band images of representative outliers from GaZNet-C9, which are not outliers anymore for GaZNet-1. In each $r$-band image, we report the spec-$z$ on the top, the GaZNet-C9 photo-$z$ on the bottom-left and the GaZNet-1 morphoto-$z$ on the bottom-right.
As a comparison, in Fig. \ref{fig:outliers}, we also show outliers from GaZNet-C9 that are still outliers for GaZNet-1. The color images in this figure give a fair idea of the galaxy SEDs, while the $r$-band images illustrate the corresponding ``morphological'' features that the GaZNet-1 uses to improve the overall predictions.

From
%the inspection of 
Fig. \ref{fig:outliers} we can distinguish four kinds of outliers for GaZNet-C9. 
The first one (A-row) is made of galaxies that are close to bright, often saturated, stars or bright large galaxies. 
%(see the three upper left panels of both raw pairs). 
In some cases, GaZNet-1 can improve the predictions and solve the discrepancy with the ground truth values (A-row). However, in some other cases, the environment is too confused to allow the CNN to guess correctly, despite 
%. However, even if 
the CNN can deblend the embedded source (E-row, see discussion below).
%, this can still be the kind of sources GaZNet-1 cannot correctly predict the redshift for, as we discuss here below.
The second type of outliers (B-row) are irregular galaxies or, generally, diffuse nearby systems. These systems are generally starforming and blue, similarly to the majority of high redshift galaxies. Thus, they typically have a higher chance to be confused with higher-$z$ systems. In this case, the GaZNet-1 can recognize the complex morphology (knots, substructures, pseudo-arms etc), or a noisier surface brightness distribution, which are typical features of closer galaxies\footnote{We can guess, here, that the CNN can learn the surface brightness fluctuation (SBF) of galaxies, which is a notorious distance indicator (see e.g., \citealt{Cantiello2005ApJ...634..239C}).}. 
The third type of outliers (C-row) is made of merging/interacting systems. For these systems, GaZNet-1 can solve the discrepancy using the information of the size of the two systems and the degree of details of the substructures, making these systems rather accurate to predict.
The fourth type of outliers (D-row) is made of blue objects, generally high-$z$ compact systems, sometimes also at low-$z$. In this case, again, the GaZNet-1 can make more accurate predictions, from the size and the round morphology. 

With this insight on the way Deep Learning help improve the predictions of photo-$z$, we can now check where it still fails. This might give us valuable indications on how we can improve the GaZNet-1 performances in future analyses.
In Fig. \ref{fig:outliers} we see three types of outliers also for the GaZNet-1.

The first one (E-row), similarly to the ones of GaZNet-C9 in the A-row above, is caused by the presence of large, bright systems. In these cases, GaZNet-1 has difficulties in either correctly deblending the source or correctly evaluating the size, especially if very compact. 
We stress here, though, that these outliers are generally fewer than all of the other kinds ($\sim 5\%$ of the total outliers for both GaZNet-1 and C9), and in the case of bright stars, can often be automatically masked out from catalogs.
The second type (F-row) is made of galaxies that have odd sizes for their redshifts, e.g., small-sized low-$z$ objects (ultra-compact galaxies? misclassified stars? etc.) or even large-sized high-$z$ systems (very massive/luminosity systems? galaxies with large diffuse haloes? etc.). As for the previous type, these systems are also quite limited in number ($\sim 4\%$), and their failure also depends on the poor training sample. In general, these outliers do not represent a significant issue.
The third type (G-row) is made of extremely compact, almost point-like and generally blue sources. These are the most abundant sample of outliers %systems 
($\sim58\%$). Although their redshift distribution is relatively sparse, they have a very similar appearance, being mainly concentrated at $z>1$ but with cases even at $z<0.5$. There is a little chance that all these systems are misclassified as stars or very compact blue galaxies (e.g., blue nuggets), although we cannot exclude some of them to be either case. The only possible option is that these are galaxies hosting active galactic nuclei (AGNs) or quasars. If so, these represent a marginal fraction of the training sample, and for this reason, they are not accurately predicted.

%There are mainly two kinds. The first kind are the ones with much bright companies. The light from the neighbours affect a lot on the central galaxies\hai{, especially for the ones with different redshifts}, hence affect the photo-$z$ determinations. However, this kind galaxies account for only a small fraction. Most of the outliers (almost $\sim2/3$) belong to the second kind. They are point-like targets, having round shapes, small sizes and high concentrated light distribution. In principle, one can get accurate photometry parameters for accurate photo-$z$ determination. However, the predictions are contrary.

%There could be two reasons. The first is that these outliers are special kind of galaxies, and account for only a very small fraction of the training samples. Therefore, the GaZNet-1 fail to capture the particular features for targets of this kind. \hai{The second reason, which is more convincing, is that these galaxies host a active galactic nuclei (AGN). 
The SED of a quasar is different from that of a typical galaxy (e.g., \citealt{2021ApJ...912...92F}), and a small fraction of quasars may not provide enough training samples. Besides, most quasars can present strong variability, since they are observed in different bands 
%images are usually observed 
at different times. This introduces fictitious color terms that increase their uncertainties in photo-$z$ measurements.
%The second reason, which is more convincing, is that these objects are not galaxies, but quasars with optical variability. Usually, different band images are observed in different times. The magnitudes of quasars will deviate from their standard SEDs because of the optical variability, hence leading to biased photo-$z$s determination. 
To verify this assumption, we check 
%the keyword of ``SG2DPHOT'', which is used to classify galaxies and point-like sources, in KiDS DR4 catalog. We find that only 61 out of the 147 outliers are classifier as galaxies, it means that the quasars account for a fraction of $\sim59$\%, which is consistent with our guess based on the visual inspection above. 
%We also checked 
the star/galaxy/quasar separation 
%based on ML based classification used 
in \cite{Khramtsov2019A&A}, based on an ML classifier.
We find that only $\sim35$\% of the 147 outliers are classified as galaxies. For the remaining $\sim65$\%, about half are classified as quasars and half as stars. Regardless of the accuracy of the ML method to classify stars and quasars, this analysis confirms that only a minority of the catastrophic events are made of galaxies, consistently 
%\rui{However, the redshifts of the ``stats" have been obtained, which means either the star/galaxy/quasar classification in \cite{Khramtsov2019A&A} or the redshift measurement from the spectra are wrong}. Overall, this is consistent 
with our guess based on the visual inspection above. In particular, the three objects in Fig. \ref{fig:outliers}-G, are all quasars in the ML classification. 

If indeed, the outliers are dominated by misclassified stars and AGNs/quasars, we can easily figure that optimizing the classification of these groups of contaminants %removed before the prediction, the 
would reduce the overall outlier fraction 
%could reduce 
down to a tiny value, $\sim0.3$\%.
%However, the redshift of these outliers have been obtained, meaning that they are not stars. Therefore, for GaZNets, $\sim 60$\% of the outliers are quasars. If these quasars are removed before the prediction, the outliers could reduce to $\sim0.3$\%, which is much better than the previous results. However, to remove all of the quasars from the galaxies is still a challenge work. A better classification method is need. We are developing a Machine Learning tools for this task by combine the information from PSFs, images and SEDs.

%In the second two rows of Fig. \ref{fig:outliers}, we show some galaxies which are outliers in the GaZNet-C, but not outliers in GaZNet. This can help to know for what kind of galaxies the GaZNet-1 can correct the worse prediction from GaZNet-C.
%We find there are mainly four kinds of targets, 1) galaxies with small size and round shape; 2) galaxies with much brighter neighbours; 3) galaxies with diffuse substructures or arms; 4) galaxies with a pair. The first two kinds are the same as that in the first two rows. It means that, compared with GaZNet-C9, the GaZNet-1 can can correct the photo-$z$ for part of these targets. For the second two kinds, almost no similar galaxies can be found to be outliers of GaZNet, which means that by introduce the images, the GaZNet-1 can fully correct the their photo-$z$s.

\subsection{Other tests}
\label{sec:more_test}
The four GaZNets illustrated in \S\ref{sec:arch} and discussed in \S\ref{sec:perform} and \S\ref{sec:discussion} have been distilled by a number of 
%Apart from the three tools we show above, we also have tested 
other models we have tested, with different kinds of inputs and different ML structures. Among these, we focus on two other experiments where we have tested two set-ups that, in principle, can 
%be responsible of significant variation of 
impact the final results. The parameters describing the performances of these two further configurations
%introduced in Sect. \ref{testing} 
are shown in Table \ref{tb:other_test}. Here below, we summarize their properties and the major results:
\begin{enumerate}
\item {\bf GaZNet-81pix to test the cutout size.} The GaZNets work on images with a size of $8''\times 8''$, which might be too small to collect the light of the whole galaxies and their environments, thus leaving some important features that the CNNs can not see.
%unseen to the CNN. 
In order to check this, we have tested the GaZNet-1 on images with twice the size of each side ($16''\times 16''$). Compared with the previous result in Table \ref{tb:other_test}, the parameters remain almost unchanged. A possible reason 
%The explanation could be 
is that the features that the CNNs extract from images are concentrated in the high SNR regions of the galaxies, while the outer regions bring little information, about both
%both about 
the galaxy properties and the environment, to improve the redshift estimates. Using these arguments, one can ask whether the standard $8''\times 8''$ cutouts are too large and one can use a smaller cutout. To check this, we also tested $4''\times 4''$ and found slightly worse results, so we kept the $8''\times 8''$ as the best choice.

%galaxy properties (e.g. structures, colors) can be well captured by the CNN even working on images with quite small sizes, and larger images will not add more useful information.
%\item {\bf GaZNet-C4 (4 band catalog):} In Section \ref{sec:testing} we have seen that the GaZNet-C, input with 9-band catalog, performs better than GaZNet-I, which is fed by only 4 band images. It is hard to know why GaZNet-C is better, it is because there are information from 5 more bands, or because the catalog are more suitable for a ML tool to capture the features for redshift determination? To check this, we train the GaZNet-C with the catalog of only 4-bands, the same as how many bands we input into GaZNet-I. From the parameters In Table \hai{Table }\ref{tb:more_test}, we conclude that, with the data of the same bands, GaZNet-I performs much better than GaZNet-C, means that images contains more information for the photz determination than catalogs of magnitudes and colors, and the CNNs can effectively capture these information.
\item {\bf GaZNet-C9I4 to test the 9-band catalogs plus 4-band images:} GaZNet-1 makes use of 9-band catalogs and only $r$-band images. To check if the addition of images on other bands can produce a sensitive improvement, we train a GaZNet with the 4 optical band images plus the 9-band catalogs. We report the statistical parameters obtained with this new GaZNet in Table \ref{tb:other_test}: generally speaking, there is no obvious improvement. Some tiny differences are compatible with the random statistical effects in the training process. Even taking these results at face values, compared to the GaZNet-1, the computing time registered by the GaZNet-C9I4 is almost 3 times longer.
%larger
For these reasons, we can discard this solution for the poor cost-benefit ratio.
\end{enumerate}
\begin{table}
\footnotesize
\begin{center}
\caption{\label{tb:other_test} Statistical properties of the predictions.}
\begin{tabular}{c c c c}
\hline \hline
CNN model & Out. fr. &$\mu_{\delta z}$& NMAD \\
 \hline
\multicolumn{4}{c}{Lower redshit galaxies}\\
 \hline
GaZNet-81pix & 0.004& 0.001& 0.015 \\
GaZNet-C9I4  & 0.004& -0.002& 0.015 \\
\hline
\multicolumn{4}{c}{higher redshift galaxies} \\
\hline
GaZNet-81pix& 0.15& -0.039& 0.045\\
GaZNet-C9I4 & 0.124& -0.033& 0.033\\
\hline \hline
\end{tabular}
\end{center}
\begin{flushleft}
\textsc{Note.} ----
Outlier fraction, mean bias and NMAD (see \S \ref{sec:Statistical_parameters}) %Statistical properties of the predictions 
from the further tools tested in \S \ref{sec:more_test}.
%Statistical properties of the prediction for more tests which are described in \ref{sec:more_test}. From left to right, we show the fraction of outliers, the mean bias and the NMAD as defined in \S \ref{sec:Statistical_parameters}.
\end{flushleft}
\end{table}
%Apart from the tests above, one can do some more tests with different inputs, such as 9-band images only, 9-band images plus 9-band catalog. However, here we will leave these tests to the future work. On one hand, to deal with the 9-band images for 100k galaxies is beyond the capability of our present hard wares. On the other hand, we have clarified that, with the same catalogs, and images from more bands,  GaZNet-C9I4 do not obviously improve the photo-$z$s with respected to GaZNet, but require much more computing resource and time.

\section{Conclusions}
Several millions of galaxies have been observed in the third generation wide-field sky surveys, and tens of billions of
%up to ten billion 
galaxies will be observed in the next ten years by the fourth generation surveys from ground and space. This enormous amount of data provides an unprecedented opportunity to study in detail the evolution of galaxies, and constrain the cosmological parameters with unprecedented accuracy. To fully conduct these studies over the expected gigantic datasets, fast and accurate photo-$z$ are indispensable.
In this work, we have explored the feasibility of determining the redshift with ML by combining the images and photometry catalogs. We designed 4 ML tools, named GaZNet-I4, GaZNet-C4 GaZNet-C9 and GaZNet-1. The inputs for these tools are 4-band images, 4-band catalogs, 9-band catalogs, and $r$-band images plus 9-band catalogs, respectively. We have trained using a sample of $\sim 140,000$ spectra from different spectroscopic surveys.
%which 
The training sample is dominated by bright (MAG$\_$AUTO$<21$) and low redshift ($z<0.8$) galaxies, which provides a quite accurate knowledge base in this parameter space.
%, was collected from different spectroscopic surveys. 
%However, even with such an unbalanced sample 
On the other hand, the higher-$z$ and fainter magnitudes are poorly covered by the training set. 
%of enough higher redshift galaxies, part of 
Despite that, we have shown that the four tools, especially GaZNet-1, still return  accurate predictions also at
%in the whole range of 
$z>0.8$.

In more details, our tests show that accurate morpho-$z$ can be directly obtained from the multi-band images ($u, g, r, i$) by GaZNet-I4, with less outliers and smaller scatters than GaZNet-C4 using only four-band optical aperture photometry.
We have also seen that the combination of optical and NIR photometry in 9-band catalogs, by GaZNet-C9, can provide a much better determination of photo-$z$. However, the information added by even one single-band high-quality images, as tested with our GaZNet-1, can achieve noticeable performances that highly overtake the ones registered for GaZNet-C9. The statistical errors are $\sim10-35$\% smaller at different redshift bins, while the outlier fraction reduces by $43$\% for lower redshift galaxies and $46\%$ for higher redshift galaxies. We have estimated the variation of the scatter as a function of the redshift, over the range of $z=0-3$, of the order of $\delta_z=0.038(1+z)$. This is heavily affected by the poor coverage of the training base at large redshifts and we expect to improve significantly this prediction by adding a few thousands more galaxies in this redshift range.
%Finally, the estimated precision (NMAD=0.014 for lower redshift galaxies and NMAD=0.041 for higher redshift galaxies), is well below the requirements for weak gravitational lensing studies in next generation ground-base surveys (e.g., NMAD=0.05 in VR/LSST, \citealt{LSST2009arXiv0912.0201L}), although this has been currently tested only on a relatively bright sample with AB magnitude $r_{auto}\leq22$ (see \ref{fig:statistics_of_bias}).}

By visually inspecting the images of all outliers produced by the GaZNet-C9 and GaZNet-1,
we have confidently demonstrated that the largest portion of the catastrophic estimates correspond to systems that are AGNs/quasars. This is corroborated by an independent ML classification from \citet{Khramtsov2019A&A}.
%we have confidently demonstrated, corroborated by an independent ML classification from \citet{Khramtsov2019A&A}, that the largest portion of the catastrophic estimates correspond to systems that are AGNs/quasars.
If these contaminants are correctly separated from galaxies, the overall outlier fraction of GaZNet-1 can reduce to 0.3\%. This is potentially an impressive result, that, combined with the rather high precision and small $\delta_z$,
%,} which we have estimated of the order of $\delta_z=0.038(1+z)$ over a redshift range of $z=0-3$. 
will make the GaZNets performance close to the requirements for galaxy evolution and cosmology studies from the 4th generation surveys (e.g.,  \citealt{Izevic+19_LSST}; \citealt{Laureijs+11_Euclid}, \citealt{Zhan+18_csst}).

\begin{acknowledgements}
Rui Li and Ran Li acknowledges the support of National Nature Science Foundation of China (Nos 11988101,11773032,12022306), the science research grants from the China Manned Space Project (No CMS-CSST-2021-B01,CMS-CSST-2021-A01) and the support from K.C.Wong Education Foundation. 
NRN acknowledge financial support from the “One hundred top talent program of Sun Yat-sen University” grant N. 71000-18841229. 
MB is supported by the Polish National Science Center through grants no. 2020/38/E/ST9/00395, 2018/30/E/ST9/00698, 2018/31/G/ST9/03388 and 2020/39/B/ST9/03494, and by the Polish Ministry of Science and Higher Education through grant DIR/WK/2018/12.
This work is based on observations made with ESO Telescopes at the La Silla Paranal Observatory under programme IDs 177.A-3016, 177.A-3017, 177.A-3018 and 179.A-2004, and on data products produced by the KiDS consortium. 
\end{acknowledgements}


\bibliographystyle{aa}
\bibliography{myrefs}

\begin{thebibliography}{96}
\expandafter\ifx\csname natexlab\endcsname\relax\def\natexlab#1{#1}\fi

\bibitem[{{Abbott} {et~al.}(2018){Abbott}, {Abdalla}, {Alarcon}, {Aleksi{\'c}},
  {Allam}, {Allen}, {Amara}, {Annis}, {Asorey}, {Avila}, {Bacon}, {Balbinot},
  {Banerji}, {Banik}, {Barkhouse}, {Baumer}, {Baxter}, {Bechtol}, {Becker},
  {Benoit-L{\'e}vy}, {Benson}, {Bernstein}, {Bertin}, {Blazek}, {Bridle},
  {Brooks}, {Brout}, {Buckley-Geer}, {Burke}, {Busha}, {Campos}, {Capozzi},
  {Carnero Rosell}, {Carrasco Kind}, {Carretero}, {Castander}, {Cawthon},
  {Chang}, {Chen}, {Childress}, {Choi}, {Conselice}, {Crittenden}, {Crocce},
  {Cunha}, {D'Andrea}, {da Costa}, {Das}, {Davis}, {Davis}, {De Vicente},
  {DePoy}, {DeRose}, {Desai}, {Diehl}, {Dietrich}, {Dodelson}, {Doel},
  {Drlica-Wagner}, {Eifler}, {Elliott}, {Elsner}, {Elvin-Poole}, {Estrada},
  {Evrard}, {Fang}, {Fernandez}, {Fert{\'e}}, {Finley}, {Flaugher}, {Fosalba},
  {Friedrich}, {Frieman}, {Garc{\'\i}a-Bellido}, {Garcia-Fernandez}, {Gatti},
  {Gaztanaga}, {Gerdes}, {Giannantonio}, {Gill}, {Glazebrook}, {Goldstein},
  {Gruen}, {Gruendl}, {Gschwend}, {Gutierrez}, {Hamilton}, {Hartley}, {Hinton},
  {Honscheid}, {Hoyle}, {Huterer}, {Jain}, {James}, {Jarvis}, {Jeltema},
  {Johnson}, {Johnson}, {Kacprzak}, {Kent}, {Kim}, {King}, {Kirk}, {Kokron},
  {Kovacs}, {Krause}, {Krawiec}, {Kremin}, {Kuehn}, {Kuhlmann}, {Kuropatkin},
  {Lacasa}, {Lahav}, {Li}, {Liddle}, {Lidman}, {Lima}, {Lin}, {MacCrann},
  {Maia}, {Makler}, {Manera}, {March}, {Marshall}, {Martini}, {McMahon},
  {Melchior}, {Menanteau}, {Miquel}, {Miranda}, {Mudd}, {Muir}, {M{\"o}ller},
  {Neilsen}, {Nichol}, {Nord}, {Nugent}, {Ogando}, {Palmese}, {Peacock},
  {Peiris}, {Peoples}, {Percival}, {Petravick}, {Plazas}, {Porredon}, {Prat},
  {Pujol}, {Rau}, {Refregier}, {Ricker}, {Roe}, {Rollins}, {Romer}, {Roodman},
  {Rosenfeld}, {Ross}, {Rozo}, {Rykoff}, {Sako}, {Salvador}, {Samuroff},
  {S{\'a}nchez}, {Sanchez}, {Santiago}, {Scarpine}, {Schindler}, {Scolnic},
  {Secco}, {Serrano}, {Sevilla-Noarbe}, {Sheldon}, {Smith}, {Smith}, {Smith},
  {Soares-Santos}, {Sobreira}, {Suchyta}, {Tarle}, {Thomas}, {Troxel},
  {Tucker}, {Tucker}, {Uddin}, {Varga}, {Vielzeuf}, {Vikram}, {Vivas},
  {Walker}, {Wang}, {Wechsler}, {Weller}, {Wester}, {Wolf}, {Yanny}, {Yuan},
  {Zenteno}, {Zhang}, {Zhang}, {Zuntz}, \& {Dark Energy Survey
  Collaboration}}]{Abbott2018PhRvD..98d3526A}
{Abbott}, T.~M.~C., {Abdalla}, F.~B., {Alarcon}, A., {et~al.} 2018, \prd, 98,
  043526

\bibitem[{{Abbott} {et~al.}(2022){Abbott}, {Aguena}, {Alarcon}, {Allam},
  {Alves}, {Amon}, {Andrade-Oliveira}, {Annis}, {Avila}, {Bacon}, {Baxter},
  {Bechtol}, {Becker}, {Bernstein}, {Bhargava}, {Birrer}, {Blazek},
  {Brandao-Souza}, {Bridle}, {Brooks}, {Buckley-Geer}, {Burke}, {Camacho},
  {Campos}, {Carnero Rosell}, {Carrasco Kind}, {Carretero}, {Castander},
  {Cawthon}, {Chang}, {Chen}, {Chen}, {Choi}, {Conselice}, {Cordero},
  {Costanzi}, {Crocce}, {da Costa}, {da Silva Pereira}, {Davis}, {Davis}, {De
  Vicente}, {DeRose}, {Desai}, {Di Valentino}, {Diehl}, {Dietrich}, {Dodelson},
  {Doel}, {Doux}, {Drlica-Wagner}, {Eckert}, {Eifler}, {Elsner}, {Elvin-Poole},
  {Everett}, {Evrard}, {Fang}, {Farahi}, {Fernandez}, {Ferrero}, {Fert{\'e}},
  {Fosalba}, {Friedrich}, {Frieman}, {Garc{\'\i}a-Bellido}, {Gatti},
  {Gaztanaga}, {Gerdes}, {Giannantonio}, {Giannini}, {Gruen}, {Gruendl},
  {Gschwend}, {Gutierrez}, {Harrison}, {Hartley}, {Herner}, {Hinton},
  {Hollowood}, {Honscheid}, {Hoyle}, {Huff}, {Huterer}, {Jain}, {James},
  {Jarvis}, {Jeffrey}, {Jeltema}, {Kovacs}, {Krause}, {Kron}, {Kuehn},
  {Kuropatkin}, {Lahav}, {Leget}, {Lemos}, {Liddle}, {Lidman}, {Lima}, {Lin},
  {MacCrann}, {Maia}, {Marshall}, {Martini}, {McCullough}, {Melchior},
  {Mena-Fern{\'a}ndez}, {Menanteau}, {Miquel}, {Mohr}, {Morgan}, {Muir},
  {Myles}, {Nadathur}, {Navarro-Alsina}, {Nichol}, {Ogando}, {Omori},
  {Palmese}, {Pandey}, {Park}, {Paz-Chinch{\'o}n}, {Petravick}, {Pieres},
  {Plazas Malag{\'o}n}, {Porredon}, {Prat}, {Raveri}, {Rodriguez-Monroy},
  {Rollins}, {Romer}, {Roodman}, {Rosenfeld}, {Ross}, {Rykoff}, {Samuroff},
  {S{\'a}nchez}, {Sanchez}, {Sanchez}, {Sanchez Cid}, {Scarpine}, {Schubnell},
  {Scolnic}, {Secco}, {Serrano}, {Sevilla-Noarbe}, {Sheldon}, {Shin}, {Smith},
  {Soares-Santos}, {Suchyta}, {Swanson}, {Tabbutt}, {Tarle}, {Thomas}, {To},
  {Troja}, {Troxel}, {Tucker}, {Tutusaus}, {Varga}, {Walker}, {Weaverdyck},
  {Wechsler}, {Weller}, {Yanny}, {Yin}, {Zhang}, {Zuntz}, \& {DES
  Collaboration}}]{Abbott2022PhRvD.105b3520A}
{Abbott}, T.~M.~C., {Aguena}, M., {Alarcon}, A., {et~al.} 2022, \prd, 105,
  023520

\bibitem[{{Abdalla} {et~al.}(2008){Abdalla}, {Amara}, {Capak}, {Cypriano},
  {Lahav}, \& {Rhodes}}]{Abdalla2008MNRAS}
{Abdalla}, F.~B., {Amara}, A., {Capak}, P., {et~al.} 2008, \mnras, 387, 969

\bibitem[{{Ackermann} {et~al.}(2018){Ackermann}, {Schawinski}, {Zhang},
  {Weigel}, \& {Turp}}]{2018MNRAS.479..415A}
{Ackermann}, S., {Schawinski}, K., {Zhang}, C., {Weigel}, A.~K., \& {Turp},
  M.~D. 2018, \mnras, 479, 415

\bibitem[{{Adhikari} {et~al.}(2021){Adhikari}, {Shin}, {Jain}, {Hilton},
  {Baxter}, {Chang}, {Wechsler}, {Battaglia}, {Bond}, {Bocquet}, {Choi},
  {DeRose}, {Devlin}, {Dunkley}, {Evrard}, {Ferraro}, {Hill}, {Hughes},
  {Gallardo}, {Lokken}, {MacInnis}, {Madhavacheril}, {McMahon}, {Nati},
  {Newburgh}, {Niemack}, {Page}, {Palmese}, {Partridge}, {Rozo}, {Rykoff},
  {Salatino}, {Schillaci}, {Sehgal}, {Sif{\'o}n}, {To}, {Wollack}, {Wu}, {Xu},
  {Aguena}, {Allam}, {Amon}, {Annis}, {Avila}, {Bacon}, {Bertin}, {Bhargava},
  {Brooks}, {Burke}, {Rosell}, {Kind}, {Carretero}, {Castander}, {Choi},
  {Costanzi}, {da Costa}, {Vicente}, {Desai}, {Diehl}, {Doel}, {Everett},
  {Ferrero}, {Fert{\'e}}, {Flaugher}, {Fosalba}, {Frieman},
  {Garc{\'\i}a-Bellido}, {Gaztanaga}, {Gruen}, {Gruendl}, {Gschwend},
  {Gutierrez}, {Hartley}, {Hinton}, {Hollowood}, {Honscheid}, {James},
  {Jeltema}, {Kuehn}, {Kuropatkin}, {Lahav}, {Lima}, {Maia}, {Marshall},
  {Martini}, {Melchior}, {Menanteau}, {Miquel}, {Morgan}, {L.~C. Ogando},
  {Paz-Chinch{\'o}n}, {Malag{\'o}n}, {Sanchez}, {Santiago}, {Scarpine},
  {Serrano}, {Sevilla-Noarbe}, {Smith}, {Soares-Santos}, {Suchyta}, {E.~C.
  Swanson}, {Varga}, {Wilkinson}, {Zhang}, {Austermann}, {Beall}, {Becker},
  {Denison}, {Duff}, {Hilton}, {Hubmayr}, {Ullom}, {Lanen}, {Vale}, {Vale}, \&
  {Vale}}]{Adhikari2021ApJ...923...37A}
{Adhikari}, S., {Shin}, T.-h., {Jain}, B., {et~al.} 2021, \apj, 923, 37

\bibitem[{{Ahumada} {et~al.}(2020){Ahumada}, {Prieto}, {Almeida}, {Anders},
  {Anderson}, {Andrews}, {Anguiano}, {Arcodia}, {Armengaud}, {Aubert}, {Avila},
  {Avila-Reese}, {Badenes}, {Balland}, {Barger}, {Barrera-Ballesteros}, {Basu},
  {Bautista}, {Beaton}, {Beers}, {Benavides}, {Bender}, {Bernardi}, {Bershady},
  {Beutler}, {Bidin}, {Bird}, {Bizyaev}, {Blanc}, {Blanton}, {Boquien},
  {Borissova}, {Bovy}, {Brandt}, {Brinkmann}, {Brownstein}, {Bundy}, {Bureau},
  {Burgasser}, {Burtin}, {Cano-D{\'\i}az}, {Capasso}, {Cappellari}, {Carrera},
  {Chabanier}, {Chaplin}, {Chapman}, {Cherinka}, {Chiappini}, {Doohyun Choi},
  {Chojnowski}, {Chung}, {Clerc}, {Coffey}, {Comerford}, {Comparat}, {da
  Costa}, {Cousinou}, {Covey}, {Crane}, {Cunha}, {Ilha}, {Dai}, {Damsted},
  {Darling}, {Davidson}, {Davies}, {Dawson}, {De}, {de la Macorra}, {De Lee},
  {Queiroz}, {Deconto Machado}, {de la Torre}, {Dell'Agli}, {du Mas des
  Bourboux}, {Diamond-Stanic}, {Dillon}, {Donor}, {Drory}, {Duckworth},
  {Dwelly}, {Ebelke}, {Eftekharzadeh}, {Davis Eigenbrot}, {Elsworth},
  {Eracleous}, {Erfanianfar}, {Escoffier}, {Fan}, {Farr},
  {Fern{\'a}ndez-Trincado}, {Feuillet}, {Finoguenov}, {Fofie},
  {Fraser-McKelvie}, {Frinchaboy}, {Fromenteau}, {Fu}, {Galbany}, {Garcia},
  {Garc{\'\i}a-Hern{\'a}ndez}, {Oehmichen}, {Ge}, {Maia}, {Geisler}, {Gelfand},
  {Goddy}, {Gonzalez-Perez}, {Grabowski}, {Green}, {Grier}, {Guo}, {Guy},
  {Harding}, {Hasselquist}, {Hawken}, {Hayes}, {Hearty}, {Hekker}, {Hogg},
  {Holtzman}, {Horta}, {Hou}, {Hsieh}, {Huber}, {Hunt}, {Chitham}, {Imig},
  {Jaber}, {Angel}, {Johnson}, {Jones}, {J{\"o}nsson}, {Jullo}, {Kim},
  {Kinemuchi}, {Kirkpatrick}, {Kite}, {Klaene}, {Kneib}, {Kollmeier}, {Kong},
  {Kounkel}, {Krishnarao}, {Lacerna}, {Lan}, {Lane}, {Law}, {Le Goff}, {Leung},
  {Lewis}, {Li}, {Lian}, {Lin}, {Long}, {Longa-Pe{\~n}a}, {Lundgren}, {Lyke},
  {Ted Mackereth}, {MacLeod}, {Majewski}, {Manchado}, {Maraston}, {Martini},
  {Masseron}, {Masters}, {Mathur}, {McDermid}, {Merloni}, {Merrifield},
  {M{\'e}sz{\'a}ros}, {Miglio}, {Minniti}, {Minsley}, {Miyaji}, {Mohammad},
  {Mosser}, {Mueller}, {Muna}, {Mu{\~n}oz-Guti{\'e}rrez}, {Myers}, {Nadathur},
  {Nair}, {Nandra}, {do Nascimento}, {Nevin}, {Newman}, {Nidever}, {Nitschelm},
  {Noterdaeme}, {O'Connell}, {Olmstead}, {Oravetz}, {Oravetz}, {Osorio},
  {Pace}, {Padilla}, {Palanque-Delabrouille}, {Palicio}, {Pan}, {Pan},
  {Parker}, {Paviot}, {Peirani}, {Ram{\'r}ez}, {Penny}, {Percival},
  {Perez-Fournon}, {P{\'e}rez-R{\`a}fols}, {Petitjean}, {Pieri},
  {Pinsonneault}, {Poovelil}, {Povick}, {Prakash}, {Price-Whelan}, {Raddick},
  {Raichoor}, {Ray}, {Rembold}, {Rezaie}, {Riffel}, {Riffel}, {Rix}, {Robin},
  {Roman-Lopes}, {Rom{\'a}n-Z{\'u}{\~n}iga}, {Rose}, {Ross}, {Rossi},
  {Rowlands}, {Rubin}, {Salvato}, {S{\'a}nchez}, {S{\'a}nchez-Menguiano},
  {S{\'a}nchez-Gallego}, {Sayres}, {Schaefer}, {Schiavon}, {Schimoia},
  {Schlafly}, {Schlegel}, {Schneider}, {Schultheis}, {Schwope}, {Seo},
  {Serenelli}, {Shafieloo}, {Shamsi}, {Shao}, {Shen}, {Shetrone}, {Shirley},
  {Aguirre}, {Simon}, {Skrutskie}, {Slosar}, {Smethurst}, {Sobeck}, {Sodi},
  {Souto}, {Stark}, {Stassun}, {Steinmetz}, {Stello}, {Stermer},
  {Storchi-Bergmann}, {Streblyanska}, {Stringfellow}, {Stutz}, {Su{\'a}rez},
  {Sun}, {Taghizadeh-Popp}, {Talbot}, {Tayar}, {Thakar}, {Theriault}, {Thomas},
  {Thomas}, {Tinker}, {Tojeiro}, {Toledo}, {Tremonti}, {Troup}, {Tuttle},
  {Unda-Sanzana}, {Valentini}, {Vargas-Gonz{\'a}lez}, {Vargas-Maga{\~n}a},
  {V{\'a}zquez-Mata}, {Vivek}, {Wake}, {Wang}, {Weaver}, {Weijmans}, {Wild},
  {Wilson}, {Wilson}, {Wolthuis}, {Wood-Vasey}, {Yan}, {Yang}, {Y{\`e}che},
  {Zamora}, {Zarrouk}, {Zasowski}, {Zhang}, {Zhao}, {Zhao}, {Zheng}, {Zheng},
  {Zhu}, \& {Zou}}]{Ahumada2020ApJS..249....3A}
{Ahumada}, R., {Prieto}, C.~A., {Almeida}, A., {et~al.} 2020, \apjs, 249, 3

\bibitem[{{Aihara} {et~al.}(2018){Aihara}, {Arimoto}, {Armstrong}, {Arnouts},
  {Bahcall}, {Bickerton}, {Bosch}, {Bundy}, {Capak}, {Chan}, {Chiba}, {Coupon},
  {Egami}, {Enoki}, {Finet}, {Fujimori}, {Fujimoto}, {Furusawa}, {Furusawa},
  {Goto}, {Goulding}, {Greco}, {Greene}, {Gunn}, {Hamana}, {Harikane},
  {Hashimoto}, {Hattori}, {Hayashi}, {Hayashi}, {He{\l}miniak}, {Higuchi},
  {Hikage}, {Ho}, {Hsieh}, {Huang}, {Huang}, {Ikeda}, {Imanishi}, {Inoue},
  {Iwasawa}, {Iwata}, {Jaelani}, {Jian}, {Kamata}, {Karoji}, {Kashikawa},
  {Katayama}, {Kawanomoto}, {Kayo}, {Koda}, {Koike}, {Kojima}, {Komiyama},
  {Konno}, {Koshida}, {Koyama}, {Kusakabe}, {Leauthaud}, {Lee}, {Lin}, {Lin},
  {Lupton}, {Mand elbaum}, {Matsuoka}, {Medezinski}, {Mineo}, {Miyama},
  {Miyatake}, {Miyazaki}, {Momose}, {More}, {More}, {Moritani}, {Moriya},
  {Morokuma}, {Mukae}, {Murata}, {Murayama}, {Nagao}, {Nakata}, {Niida},
  {Niikura}, {Nishizawa}, {Obuchi}, {Oguri}, {Oishi}, {Okabe}, {Okamoto},
  {Okura}, {Ono}, {Onodera}, {Onoue}, {Osato}, {Ouchi}, {Price}, {Pyo}, {Sako},
  {Sawicki}, {Shibuya}, {Shimasaku}, {Shimono}, {Shirasaki}, {Silverman},
  {Simet}, {Speagle}, {Spergel}, {Strauss}, {Sugahara}, {Sugiyama}, {Suto},
  {Suyu}, {Suzuki}, {Tait}, {Takada}, {Takata}, {Tamura}, {Tanaka}, {Tanaka},
  {Tanaka}, {Tanaka}, {Terai}, {Terashima}, {Toba}, {Tominaga}, {Toshikawa},
  {Turner}, {Uchida}, {Uchiyama}, {Umetsu}, {Uraguchi}, {Urata}, {Usuda},
  {Utsumi}, {Wang}, {Wang}, {Wong}, {Yabe}, {Yamada}, {Yamanoi}, {Yasuda},
  {Yeh}, {Yonehara}, \& {Yuma}}]{Aihara+18_HSC}
{Aihara}, H., {Arimoto}, N., {Armstrong}, R., {et~al.} 2018, \pasj, 70, S4

\bibitem[{{Amaro} {et~al.}(2021){Amaro}, {Cavuoti}, {Brescia}, {Riccio},
  {Tortora}, {Delli Veneri}, {Napolitano}, {Radovich}, \&
  {Longo}}]{Amaro2021+photz}
{Amaro}, V., {Cavuoti}, S., {Brescia}, M., {et~al.} 2021, {Rejection Criteria
  Based on Outliers in the KiDS Photometric Redshifts and PDF Distributions
  Derived by Machine Learning}, ed. I.~{Zelinka}, M.~{Brescia}, \& D.~{Baron},
  Vol.~39, 245--264

\bibitem[{{Baldry} {et~al.}(2018{\natexlab{a}}){Baldry}, {Liske}, {Brown},
  {Robotham}, {Driver}, {Dunne}, {Alpaslan}, {Brough}, {Cluver}, {Eardley},
  {Farrow}, {Heymans}, {Hildebrandt}, {Hopkins}, {Kelvin}, {Loveday},
  {Moffett}, {Norberg}, {Owers}, {Taylor}, {Wright}, {Bamford},
  {Bland-Hawthorn}, {Bourne}, {Bremer}, {Colless}, {Conselice}, {Croom},
  {Davies}, {Foster}, {Grootes}, {Holwerda}, {Jones}, {Kafle}, {Kuijken},
  {Lara-Lopez}, {L{\'o}pez-S{\'a}nchez}, {Meyer}, {Phillipps}, {Sutherland},
  {van Kampen}, \& {Wilkins}}]{Baldry2018MNRAS.474.3875B}
{Baldry}, I.~K., {Liske}, J., {Brown}, M.~J.~I., {et~al.} 2018{\natexlab{a}},
  \mnras, 474, 3875

\bibitem[{{Baldry} {et~al.}(2018{\natexlab{b}}){Baldry}, {Liske}, {Brown},
  {Robotham}, {Driver}, {Dunne}, {Alpaslan}, {Brough}, {Cluver}, {Eardley},
  {Farrow}, {Heymans}, {Hildebrandt}, {Hopkins}, {Kelvin}, {Loveday},
  {Moffett}, {Norberg}, {Owers}, {Taylor}, {Wright}, {Bamford},
  {Bland-Hawthorn}, {Bourne}, {Bremer}, {Colless}, {Conselice}, {Croom},
  {Davies}, {Foster}, {Grootes}, {Holwerda}, {Jones}, {Kafle}, {Kuijken},
  {Lara-Lopez}, {L{\'o}pez-S{\'a}nchez}, {Meyer}, {Phillipps}, {Sutherland},
  {van Kampen}, \& {Wilkins}}]{Baldry2018MNRAS_GAMADR3}
{Baldry}, I.~K., {Liske}, J., {Brown}, M.~J.~I., {et~al.} 2018{\natexlab{b}},
  \mnras, 474, 3875

\bibitem[{{Banerji} {et~al.}(2008){Banerji}, {Abdalla}, {Lahav}, \&
  {Lin}}]{Banerji2008MNRAS}
{Banerji}, M., {Abdalla}, F.~B., {Lahav}, O., \& {Lin}, H. 2008, \mnras, 386,
  1219

\bibitem[{{Baum}(1962)}]{1962IAUS...15..390B}
{Baum}, W.~A. 1962, in Problems of Extra-Galactic Research, ed. G.~C.
  {McVittie}, Vol.~15, 390

\bibitem[{{Behroozi} {et~al.}(2019){Behroozi}, {Wechsler}, {Hearin}, \&
  {Conroy}}]{Behroozi2019MNRAS.488.3143B}
{Behroozi}, P., {Wechsler}, R.~H., {Hearin}, A.~P., \& {Conroy}, C. 2019,
  \mnras, 488, 3143

\bibitem[{{Ben{\'\i}tez}(2000)}]{2000ApJ...536..571B_BPZ}
{Ben{\'\i}tez}, N. 2000, \apj, 536, 571

\bibitem[{{Bertin} \& {Arnouts}(1996)}]{Bertin_Arnouts96_SEx}
{Bertin}, E. \& {Arnouts}, S. 1996, \aaps, 117, 393

\bibitem[{{Bilicki} {et~al.}(2021){Bilicki}, {Dvornik}, {Hoekstra}, {Wright},
  {Chisari}, {Vakili}, {Asgari}, {Giblin}, {Heymans}, {Hildebrandt},
  {Holwerda}, {Hopkins}, {Johnston}, {Kannawadi}, {Kuijken}, {Nakoneczny},
  {Shan}, {Sonnenfeld}, \& {Valentijn}}]{Bilicki2021A&A...653A..82B}
{Bilicki}, M., {Dvornik}, A., {Hoekstra}, H., {et~al.} 2021, \aap, 653, A82

\bibitem[{{Bilicki} {et~al.}(2018){Bilicki}, {Hoekstra}, {Brown}, {Amaro},
  {Blake}, {Cavuoti}, {de Jong}, {Georgiou}, {Hildebrandt}, {Wolf}, {Amon},
  {Brescia}, {Brough}, {Costa-Duarte}, {Erben}, {Glazebrook}, {Grado},
  {Heymans}, {Jarrett}, {Joudaki}, {Kuijken}, {Longo}, {Napolitano},
  {Parkinson}, {Vellucci}, {Verdoes Kleijn}, \&
  {Wang}}]{Bilicki2018A&A...616A..69B}
{Bilicki}, M., {Hoekstra}, H., {Brown}, M.~J.~I., {et~al.} 2018, \aap, 616, A69

\bibitem[{{Bolzonella} {et~al.}(2000){Bolzonella}, {Miralles}, \&
  {Pell{\'o}}}]{Bolzonella2000A&A}
{Bolzonella}, M., {Miralles}, J.~M., \& {Pell{\'o}}, R. 2000, \aap, 363, 476

\bibitem[{{Brescia} {et~al.}(2014){Brescia}, {Cavuoti}, {Longo}, \& {De
  Stefano}}]{Brescia2014A&A}
{Brescia}, M., {Cavuoti}, S., {Longo}, G., \& {De Stefano}, V. 2014, \aap, 568,
  A126

\bibitem[{{Cantiello} {et~al.}(2005){Cantiello}, {Blakeslee}, {Raimondo},
  {Mei}, {Brocato}, \& {Capaccioli}}]{Cantiello2005ApJ...634..239C}
{Cantiello}, M., {Blakeslee}, J.~P., {Raimondo}, G., {et~al.} 2005, \apj, 634,
  239

\bibitem[{{Capaccioli} \& {Schipani}(2011)}]{Capaccioli2011Msngr.146....2C}
{Capaccioli}, M. \& {Schipani}, P. 2011, The Messenger, 146, 2

\bibitem[{{Cavuoti} {et~al.}(2012){Cavuoti}, {Brescia}, {Longo}, \&
  {Mercurio}}]{Cavuoti2012A&A_MLPQNA}
{Cavuoti}, S., {Brescia}, M., {Longo}, G., \& {Mercurio}, A. 2012, \aap, 546,
  A13

\bibitem[{{Cavuoti} {et~al.}(2015){Cavuoti}, {Brescia}, {Tortora}, {Longo},
  {Napolitano}, {Radovich}, {Barbera}, {Capaccioli}, {de Jong}, {Getman},
  {Grado}, \& {Paolillo}}]{Cavuoti+15_KIDS_I}
{Cavuoti}, S., {Brescia}, M., {Tortora}, C., {et~al.} 2015, \mnras, 452, 3100

\bibitem[{{{\'C}iprijanovi{\'c}} {et~al.}(2020){{\'C}iprijanovi{\'c}},
  {Snyder}, {Nord}, \& {Peek}}]{2020A&C....3200390C}
{{\'C}iprijanovi{\'c}}, A., {Snyder}, G.~F., {Nord}, B., \& {Peek}, J.~E.~G.
  2020, Astronomy and Computing, 32, 100390

\bibitem[{{Collister} {et~al.}(2007){Collister}, {Lahav}, {Blake}, {Cannon},
  {Croom}, {Drinkwater}, {Edge}, {Eisenstein}, {Loveday}, {Nichol}, {Pimbblet},
  {de Propris}, {Roseboom}, {Ross}, {Schneider}, {Shanks}, \&
  {Wake}}]{Collister2007MNRAS}
{Collister}, A., {Lahav}, O., {Blake}, C., {et~al.} 2007, \mnras, 375, 68

\bibitem[{{Collister} \& {Lahav}(2004)}]{Collister2004PASP..116..345C}
{Collister}, A.~A. \& {Lahav}, O. 2004, \pasp, 116, 345

\bibitem[{{Connolly}(1997)}]{Connolly1997hst}
{Connolly}, A. 1997, {The Properties of High Redshift Galaxies : A
  Near-Infrared Redshift Survey at 1<z<2}, HST Proposal

\bibitem[{{Connolly} {et~al.}(1995){Connolly}, {Csabai}, {Szalay}, {Koo},
  {Kron}, \& {Munn}}]{Connolly1995AJ}
{Connolly}, A.~J., {Csabai}, I., {Szalay}, A.~S., {et~al.} 1995, \aj, 110, 2655

\bibitem[{{Couch} {et~al.}(1983){Couch}, {Ellis}, {Godwin}, \&
  {Carter}}]{Couch1983}
{Couch}, W.~J., {Ellis}, R.~S., {Godwin}, J., \& {Carter}, D. 1983, \mnras,
  205, 1287

\bibitem[{Cun {et~al.}(1990)Cun, Boser, Denker, Henderson, \&
  Jackel}]{1990Handwritten}
Cun, Y.~L., Boser, B., Denker, J.~S., Henderson, D., \& Jackel, L.~D. 1990,
  Advances in neural information processing systems, 2, 396

\bibitem[{{de Jong} {et~al.}(2017){de Jong}, {Verdoes Kleijn}, {Erben},
  {Hildebrandt}, {Kuijken}, {Sikkema}, {Brescia}, {Bilicki}, {Napolitano},
  {Amaro}, {Begeman}, {Boxhoorn}, {Buddelmeijer}, {Cavuoti}, {Getman}, {Grado},
  {Helmich}, {Huang}, {Irisarri}, {La Barbera}, {Longo}, {McFarland},
  {Nakajima}, {Paolillo}, {Puddu}, {Radovich}, {Rifatto}, {Tortora},
  {Valentijn}, {Vellucci}, {Vriend}, {Amon}, {Blake}, {Choi}, {Conti}, {Gwyn},
  {Herbonnet}, {Heymans}, {Hoekstra}, {Klaes}, {Merten}, {Miller}, {Schneider},
  \& {Viola}}]{deJong2017A&A...604A.134D}
{de Jong}, J. T.~A., {Verdoes Kleijn}, G.~A., {Erben}, T., {et~al.} 2017, \aap,
  604, A134

\bibitem[{{de Jong} {et~al.}(2013){de Jong}, {Verdoes Kleijn}, {Kuijken}, \&
  {Valentijn}}]{deJong2013ExA....35...25D}
{de Jong}, J. T.~A., {Verdoes Kleijn}, G.~A., {Kuijken}, K.~H., \& {Valentijn},
  E.~A. 2013, Experimental Astronomy, 35, 25

\bibitem[{{de Jong} {et~al.}(2019){de Jong}, {Agertz}, {Berbel}, {Aird},
  {Alexander}, {Amarsi}, {Anders}, {Andrae}, {Ansarinejad}, {Ansorge},
  {Antilogus}, {Anwand-Heerwart}, {Arentsen}, {Arnadottir}, {Asplund}, {Auger},
  {Azais}, {Baade}, {Baker}, {Baker}, {Balbinot}, {Baldry}, {Banerji},
  {Barden}, {Barklem}, {Barth{\'e}l{\'e}my-Mazot}, {Battistini}, {Bauer},
  {Bell}, {Bellido-Tirado}, {Bellstedt}, {Belokurov}, {Bensby}, {Bergemann},
  {Bestenlehner}, {Bielby}, {Bilicki}, {Blake}, {Bland-Hawthorn}, {Boeche},
  {Boland}, {Boller}, {Bongard}, {Bongiorno}, {Bonifacio}, {Boudon}, {Brooks},
  {Brown}, {Brown}, {Br{\"u}ggen}, {Brynnel}, {Brzeski}, {Buchert},
  {Buschkamp}, {Caffau}, {Caillier}, {Carrick}, {Casagrande}, {Case}, {Casey},
  {Cesarini}, {Cescutti}, {Chapuis}, {Chiappini}, {Childress}, {Christlieb},
  {Church}, {Cioni}, {Cluver}, {Colless}, {Collett}, {Comparat}, {Cooper},
  {Couch}, {Courbin}, {Croom}, {Croton}, {Daguis{\'e}}, {Dalton}, {Davies},
  {Davis}, {de Laverny}, {Deason}, {Dionies}, {Disseau}, {Doel}, {D{\"o}scher},
  {Driver}, {Dwelly}, {Eckert}, {Edge}, {Edvardsson}, {Youssoufi}, {Elhaddad},
  {Enke}, {Erfanianfar}, {Farrell}, {Fechner}, {Feiz}, {Feltzing}, {Ferreras},
  {Feuerstein}, {Feuillet}, {Finoguenov}, {Ford}, {Fotopoulou}, {Fouesneau},
  {Frenk}, {Frey}, {Gaessler}, {Geier}, {Gentile Fusillo}, {Gerhard},
  {Giannantonio}, {Giannone}, {Gibson}, {Gillingham},
  {Gonz{\'a}lez-Fern{\'a}ndez}, {Gonzalez-Solares}, {Gottloeber}, {Gould},
  {Grebel}, {Gueguen}, {Guiglion}, {Haehnelt}, {Hahn}, {Hansen}, {Hartman},
  {Hauptner}, {Hawkins}, {Haynes}, {Haynes}, {Heiter}, {Helmi}, {Aguayo},
  {Hewett}, {Hinton}, {Hobbs}, {Hoenig}, {Hofman}, {Hook}, {Hopgood},
  {Hopkins}, {Hourihane}, {Howes}, {Howlett}, {Huet}, {Irwin}, {Iwert},
  {Jablonka}, {Jahn}, {Jahnke}, {Jarno}, {Jin}, {Jofre}, {Johl}, {Jones},
  {J{\"o}nsson}, {Jordan}, {Karovicova}, {Khalatyan}, {Kelz}, {Kennicutt},
  {King}, {Kitaura}, {Klar}, {Klauser}, {Kneib}, {Koch}, {Koposov},
  {Kordopatis}, {Korn}, {Kosmalski}, {Kotak}, {Kovalev}, {Kreckel}, {Kripak},
  {Krumpe}, {Kuijken}, {Kunder}, {Kushniruk}, {Lam}, {Lamer}, {Laurent},
  {Lawrence}, {Lehmitz}, {Lemasle}, {Lewis}, {Li}, {Lidman}, {Lind}, {Liske},
  {Lizon}, {Loveday}, {Ludwig}, {McDermid}, {Maguire}, {Mainieri}, {Mali},
  {Mandel}, {Mandel}, {Mannering}, {Martell}, {Martinez Delgado}, {Matijevic},
  {McGregor}, {McMahon}, {McMillan}, {Mena}, {Merloni}, {Meyer}, {Michel},
  {Micheva}, {Migniau}, {Minchev}, {Monari}, {Muller}, {Murphy},
  {Muthukrishna}, {Nandra}, {Navarro}, {Ness}, {Nichani}, {Nichol}, {Nicklas},
  {Niederhofer}, {Norberg}, {Obreschkow}, {Oliver}, {Owers}, {Pai},
  {Pankratow}, {Parkinson}, {Paschke}, {Paterson}, {Pecontal}, {Parry},
  {Phillips}, {Pillepich}, {Pinard}, {Pirard}, {Piskunov}, {Plank},
  {Pl{\"u}schke}, {Pons}, {Popesso}, {Power}, {Pragt}, {Pramskiy}, {Pryer},
  {Quattri}, {Queiroz}, {Quirrenbach}, {Rahurkar}, {Raichoor}, {Ramstedt},
  {Rau}, {Recio-Blanco}, {Reiss}, {Renaud}, {Revaz}, {Rhode}, {Richard},
  {Richter}, {Rix}, {Robotham}, {Roelfsema}, {Romaniello}, {Rosario},
  {Rothmaier}, {Roukema}, {Ruchti}, {Rupprecht}, {Rybizki}, {Ryde}, {Saar},
  {Sadler}, {Sahl{\'e}n}, {Salvato}, {Sassolas}, {Saunders}, {Saviauk},
  {Sbordone}, {Schmidt}, {Schnurr}, {Scholz}, {Schwope}, {Seifert}, {Shanks},
  {Sheinis}, {Sivov}, {Sk{\'u}lad{\'o}ttir}, {Smartt}, {Smedley}, {Smith},
  {Smith}, {Sorce}, {Spitler}, {Starkenburg}, {Steinmetz}, {Stilz}, {Storm},
  {Sullivan}, {Sutherland}, {Swann}, {Tamone}, {Taylor}, {Teillon}, {Tempel},
  {ter Horst}, {Thi}, {Tolstoy}, {Trager}, {Traven}, {Tremblay}, {Tresse},
  {Valentini}, {van de Weygaert}, {van den Ancker}, {Veljanoski}, {Venkatesan},
  {Wagner}, {Wagner}, {Walcher}, {Waller}, {Walton}, {Wang}, {Winkler},
  {Wisotzki}, {Worley}, {Worseck}, {Xiang}, {Xu}, {Yong}, {Zhao}, {Zheng},
  {Zscheyge}, \& {Zucker}}]{deJong2019_4MOST}
{de Jong}, R.~S., {Agertz}, O., {Berbel}, A.~A., {et~al.} 2019, The Messenger,
  175, 3

\bibitem[{{DESI Collaboration} {et~al.}(2016){DESI Collaboration}, {Aghamousa},
  {Aguilar}, {Ahlen}, {Alam}, {Allen}, {Allende Prieto}, {Annis}, {Bailey},
  {Balland}, {Ballester}, {Baltay}, {Beaufore}, {Bebek}, {Beers}, {Bell},
  {Bernal}, {Besuner}, {Beutler}, {Blake}, {Bleuler}, {Blomqvist}, {Blum},
  {Bolton}, {Briceno}, {Brooks}, {Brownstein}, {Buckley-Geer}, {Burden},
  {Burtin}, {Busca}, {Cahn}, {Cai}, {Cardiel-Sas}, {Carlberg}, {Carton},
  {Casas}, {Castander}, {Cervantes-Cota}, {Claybaugh}, {Close}, {Coker},
  {Cole}, {Comparat}, {Cooper}, {Cousinou}, {Crocce}, {Cuby}, {Cunningham},
  {Davis}, {Dawson}, {de la Macorra}, {De Vicente}, {Delubac}, {Derwent},
  {Dey}, {Dhungana}, {Ding}, {Doel}, {Duan}, {Ealet}, {Edelstein},
  {Eftekharzadeh}, {Eisenstein}, {Elliott}, {Escoffier}, {Evatt}, {Fagrelius},
  {Fan}, {Fanning}, {Farahi}, {Farihi}, {Favole}, {Feng}, {Fernandez},
  {Findlay}, {Finkbeiner}, {Fitzpatrick}, {Flaugher}, {Flender}, {Font-Ribera},
  {Forero-Romero}, {Fosalba}, {Frenk}, {Fumagalli}, {Gaensicke}, {Gallo},
  {Garcia-Bellido}, {Gaztanaga}, {Pietro Gentile Fusillo}, {Gerard},
  {Gershkovich}, {Giannantonio}, {Gillet}, {Gonzalez-de-Rivera},
  {Gonzalez-Perez}, {Gott}, {Graur}, {Gutierrez}, {Guy}, {Habib}, {Heetderks},
  {Heetderks}, {Heitmann}, {Hellwing}, {Herrera}, {Ho}, {Holland}, {Honscheid},
  {Huff}, {Hutchinson}, {Huterer}, {Hwang}, {Illa Laguna}, {Ishikawa},
  {Jacobs}, {Jeffrey}, {Jelinsky}, {Jennings}, {Jiang}, {Jimenez}, {Johnson},
  {Joyce}, {Jullo}, {Juneau}, {Kama}, {Karcher}, {Karkar}, {Kehoe}, {Kennamer},
  {Kent}, {Kilbinger}, {Kim}, {Kirkby}, {Kisner}, {Kitanidis}, {Kneib},
  {Koposov}, {Kovacs}, {Koyama}, {Kremin}, {Kron}, {Kronig}, {Kueter-Young},
  {Lacey}, {Lafever}, {Lahav}, {Lambert}, {Lampton}, {Landriau}, {Lang},
  {Lauer}, {Le Goff}, {Le Guillou}, {Le Van Suu}, {Lee}, {Lee}, {Leitner},
  {Lesser}, {Levi}, {L'Huillier}, {Li}, {Liang}, {Lin}, {Linder}, {Loebman},
  {Luki{\'c}}, {Ma}, {MacCrann}, {Magneville}, {Makarem}, {Manera}, {Manser},
  {Marshall}, {Martini}, {Massey}, {Matheson}, {McCauley}, {McDonald},
  {McGreer}, {Meisner}, {Metcalfe}, {Miller}, {Miquel}, {Moustakas}, {Myers},
  {Naik}, {Newman}, {Nichol}, {Nicola}, {Nicolati da Costa}, {Nie}, {Niz},
  {Norberg}, {Nord}, {Norman}, {Nugent}, {O'Brien}, {Oh}, {Olsen}, {Padilla},
  {Padmanabhan}, {Padmanabhan}, {Palanque-Delabrouille}, {Palmese},
  {Pappalardo}, {P{\^a}ris}, {Park}, {Patej}, {Peacock}, {Peiris}, {Peng},
  {Percival}, {Perruchot}, {Pieri}, {Pogge}, {Pollack}, {Poppett}, {Prada},
  {Prakash}, {Probst}, {Rabinowitz}, {Raichoor}, {Ree}, {Refregier}, {Regal},
  {Reid}, {Reil}, {Rezaie}, {Rockosi}, {Roe}, {Ronayette}, {Roodman}, {Ross},
  {Ross}, {Rossi}, {Rozo}, {Ruhlmann-Kleider}, {Rykoff}, {Sabiu}, {Samushia},
  {Sanchez}, {Sanchez}, {Schlegel}, {Schneider}, {Schubnell}, {Secroun},
  {Seljak}, {Seo}, {Serrano}, {Shafieloo}, {Shan}, {Sharples}, {Sholl},
  {Shourt}, {Silber}, {Silva}, {Sirk}, {Slosar}, {Smith}, {Smoot}, {Som},
  {Song}, {Sprayberry}, {Staten}, {Stefanik}, {Tarle}, {Sien Tie}, {Tinker},
  {Tojeiro}, {Valdes}, {Valenzuela}, {Valluri}, {Vargas-Magana}, {Verde},
  {Walker}, {Wang}, {Wang}, {Weaver}, {Weaverdyck}, {Wechsler}, {Weinberg},
  {White}, {Yang}, {Yeche}, {Zhang}, {Zhao}, {Zheng}, {Zhou}, {Zhou}, {Zhu},
  {Zou}, \& {Zu}}]{DESI_Collaboration_2016}
{DESI Collaboration}, {Aghamousa}, A., {Aguilar}, J., {et~al.} 2016, arXiv
  e-prints, arXiv:1611.00036

\bibitem[{{Dey} {et~al.}(2021){Dey}, {Andrews}, {Newman}, {Mao}, {Rau}, \&
  {Zhou}}]{Dey2021arXiv211203939D}
{Dey}, B., {Andrews}, B.~H., {Newman}, J.~A., {et~al.} 2021, arXiv e-prints,
  arXiv:2112.03939

\bibitem[{{Di Valentino} {et~al.}(2021){Di Valentino}, {Anchordoqui}, {Akarsu},
  {Ali-Haimoud}, {Amendola}, {Arendse}, {Asgari}, {Ballardini}, {Basilakos},
  {Battistelli}, {Benetti}, {Birrer}, {Bouchet}, {Bruni}, {Calabrese},
  {Camarena}, {Capozziello}, {Chen}, {Chluba}, {Chudaykin}, {Colg{\'a}in},
  {Cyr-Racine}, {de Bernardis}, {de Cruz P{\'e}rez}, {Delabrouille}, {Dunkley},
  {Escamilla-Rivera}, {Fert{\'e}}, {Finelli}, {Freedman}, {Frusciante},
  {Giusarma}, {G{\'o}mez-Valent}, {Handley}, {Harrison}, {Hart}, {Heavens},
  {Hildebrandt}, {Holz}, {Huterer}, {Ivanov}, {Joudaki}, {Kamionkowski},
  {Karwal}, {Knox}, {Kumar}, {Lamagna}, {Lesgourgues}, {Lucca}, {Marra},
  {Masi}, {Matarrese}, {Mazumdar}, {Melchiorri}, {Mena}, {Mersini-Houghton},
  {Miranda}, {Moreno-Pulido}, {Mota}, {Muir}, {Mukherjee}, {Niedermann},
  {Notari}, {Nunes}, {Pace}, {Paliathanasis}, {Palmese}, {Pan}, {Paoletti},
  {Pettorino}, {Piacentini}, {Poulin}, {Raveri}, {Riess}, {Salzano},
  {Saridakis}, {Sen}, {Shafieloo}, {Shajib}, {Silk}, {Silvestri}, {Sloth},
  {Smith}, {Sol{\`a} Peracaula}, {van de Bruck}, {Verde}, {Visinelli},
  {Wandelt}, {Wang}, {Wang}, {Yadav}, \& {Yang}}]{2021APh...13102606D}
{Di Valentino}, E., {Anchordoqui}, L.~A., {Akarsu}, {\"O}., {et~al.} 2021,
  Astroparticle Physics, 131, 102606

\bibitem[{{D'Isanto} {et~al.}(2018){D'Isanto}, {Cavuoti}, {Gieseke}, \&
  {Polsterer}}]{2018A&A...616A..97D}
{D'Isanto}, A., {Cavuoti}, S., {Gieseke}, F., \& {Polsterer}, K.~L. 2018, \aap,
  616, A97

\bibitem[{{D'Isanto} \& {Polsterer}(2018)}]{D'Isanto2018A&A...609A.111D}
{D'Isanto}, A. \& {Polsterer}, K.~L. 2018, \aap, 609, A111

\bibitem[{{Dom{\'\i}nguez S{\'a}nchez} {et~al.}(2018){Dom{\'\i}nguez
  S{\'a}nchez}, {Huertas-Company}, {Bernardi}, {Tuccillo}, \&
  {Fischer}}]{2018MNRAS.476.3661D}
{Dom{\'\i}nguez S{\'a}nchez}, H., {Huertas-Company}, M., {Bernardi}, M.,
  {Tuccillo}, D., \& {Fischer}, J.~L. 2018, \mnras, 476, 3661

\bibitem[{{Driver} {et~al.}(2011){Driver}, {Hill}, {Kelvin}, {Robotham},
  {Liske}, {Norberg}, {Baldry}, {Bamford}, {Hopkins}, {Loveday}, {Peacock},
  {Andrae}, {Bland-Hawthorn}, {Brough}, {Brown}, {Cameron}, {Ching}, {Colless},
  {Conselice}, {Croom}, {Cross}, {de Propris}, {Dye}, {Drinkwater}, {Ellis},
  {Graham}, {Grootes}, {Gunawardhana}, {Jones}, {van Kampen}, {Maraston},
  {Nichol}, {Parkinson}, {Phillipps}, {Pimbblet}, {Popescu}, {Prescott},
  {Roseboom}, {Sadler}, {Sansom}, {Sharp}, {Smith}, {Taylor}, {Thomas},
  {Tuffs}, {Wijesinghe}, {Dunne}, {Frenk}, {Jarvis}, {Madore}, {Meyer},
  {Seibert}, {Staveley-Smith}, {Sutherland}, \&
  {Warren}}]{Driver2011MNRAS_GAMA}
{Driver}, S.~P., {Hill}, D.~T., {Kelvin}, L.~S., {et~al.} 2011, \mnras, 413,
  971

\bibitem[{{Edge} {et~al.}(2013){Edge}, {Sutherland}, {Kuijken}, {Driver},
  {McMahon}, {Eales}, \& {Emerson}}]{Edge2013Msngr_VIKING}
{Edge}, A., {Sutherland}, W., {Kuijken}, K., {et~al.} 2013, The Messenger, 154,
  32

\bibitem[{{Edge} {et~al.}(2014){Edge}, {Sutherland}, \& {The Viking
  Team}}]{Edge+14_VIKING-DR1}
{Edge}, A., {Sutherland}, W., \& {The Viking Team}. 2014, VizieR Online Data
  Catalog, 2329, 0

\bibitem[{{Feng} {et~al.}(2021){Feng}, {Liu}, {Bai}, {Yang}, {Hu}, {Li},
  {Yang}, {Lu}, \& {Xiao}}]{2021ApJ...912...92F}
{Feng}, H.-C., {Liu}, H.~T., {Bai}, J.~M., {et~al.} 2021, \apj, 912, 92

\bibitem[{Friedman(1999)}]{Friedman99+huberloss}
Friedman, J.~H. 1999, Annals of Statistics, 29, 1189

\bibitem[{{Gal} \& {Ghahramani}(2015)}]{Gal2015arXiv150602158G}
{Gal}, Y. \& {Ghahramani}, Z. 2015, arXiv e-prints, arXiv:1506.02158

\bibitem[{{Gong} {et~al.}(2019){Gong}, {Liu}, {Cao}, {Chen}, {Fan}, {Li}, {Li},
  {Li}, {Zhang}, \& {Zhan}}]{Gong2019ApJ...883..203G}
{Gong}, Y., {Liu}, X., {Cao}, Y., {et~al.} 2019, \apj, 883, 203

\bibitem[{{Goulding} {et~al.}(2018){Goulding}, {Greene}, {Bezanson}, {Greco},
  {Johnson}, {Leauthaud}, {Matsuoka}, {Medezinski}, \&
  {Price-Whelan}}]{Goulding2018PASJ...70S..37G}
{Goulding}, A.~D., {Greene}, J.~E., {Bezanson}, R., {et~al.} 2018, \pasj, 70,
  S37

\bibitem[{{Greco} {et~al.}(2018){Greco}, {Greene}, {Strauss}, {Macarthur},
  {Flowers}, {Goulding}, {Huang}, {Kim}, {Komiyama}, {Leauthaud}, {Leisman},
  {Lupton}, {Sif{\'o}n}, \& {Wang}}]{Greco2018ApJ...857..104G}
{Greco}, J.~P., {Greene}, J.~E., {Strauss}, M.~A., {et~al.} 2018, \apj, 857,
  104

\bibitem[{{Heymans} {et~al.}(2021){Heymans}, {Tr{\"o}ster}, {Asgari}, {Blake},
  {Hildebrandt}, {Joachimi}, {Kuijken}, {Lin}, {S{\'a}nchez}, {van den Busch},
  {Wright}, {Amon}, {Bilicki}, {de Jong}, {Crocce}, {Dvornik}, {Erben},
  {Fortuna}, {Getman}, {Giblin}, {Glazebrook}, {Hoekstra}, {Joudaki},
  {Kannawadi}, {K{\"o}hlinger}, {Lidman}, {Miller}, {Napolitano}, {Parkinson},
  {Schneider}, {Shan}, {Valentijn}, {Verdoes Kleijn}, \&
  {Wolf}}]{Heymans2021A&A...646A.140H}
{Heymans}, C., {Tr{\"o}ster}, T., {Asgari}, M., {et~al.} 2021, \aap, 646, A140

\bibitem[{{Hikage} {et~al.}(2019){Hikage}, {Oguri}, {Hamana}, {More},
  {Mandelbaum}, {Takada}, {K{\"o}hlinger}, {Miyatake}, {Nishizawa}, {Aihara},
  {Armstrong}, {Bosch}, {Coupon}, {Ducout}, {Ho}, {Hsieh}, {Komiyama},
  {Lanusse}, {Leauthaud}, {Lupton}, {Medezinski}, {Mineo}, {Miyama},
  {Miyazaki}, {Murata}, {Murayama}, {Shirasaki}, {Sif{\'o}n}, {Simet},
  {Speagle}, {Spergel}, {Strauss}, {Sugiyama}, {Tanaka}, {Utsumi}, {Wang}, \&
  {Yamada}}]{Hikage2019PASJ...71...43H}
{Hikage}, C., {Oguri}, M., {Hamana}, T., {et~al.} 2019, \pasj, 71, 43

\bibitem[{{Hildebrandt} {et~al.}(2010){Hildebrandt}, {Arnouts}, {Capak},
  {Moustakas}, {Wolf}, {Abdalla}, {Assef}, {Banerji}, {Ben{\'\i}tez},
  {Brammer}, {Budav{\'a}ri}, {Carliles}, {Coe}, {Dahlen}, {Feldmann}, {Gerdes},
  {Gillis}, {Ilbert}, {Kotulla}, {Lahav}, {Li}, {Miralles}, {Purger},
  {Schmidt}, \& {Singal}}]{Hildebrandt2010A&A}
{Hildebrandt}, H., {Arnouts}, S., {Capak}, P., {et~al.} 2010, \aap, 523, A31

\bibitem[{{Hildebrandt} {et~al.}(2017){Hildebrandt}, {Viola}, {Heymans},
  {Joudaki}, {Kuijken}, {Blake}, {Erben}, {Joachimi}, {Klaes}, {Miller},
  {Morrison}, {Nakajima}, {Verdoes Kleijn}, {Amon}, {Choi}, {Covone}, {de
  Jong}, {Dvornik}, {Fenech Conti}, {Grado}, {Harnois-D{\'e}raps}, {Herbonnet},
  {Hoekstra}, {K{\"o}hlinger}, {McFarland}, {Mead}, {Merten}, {Napolitano},
  {Peacock}, {Radovich}, {Schneider}, {Simon}, {Valentijn}, {van den Busch},
  {van Uitert}, \& {Van Waerbeke}}]{Hildebrandt2017MNRAS.465.1454H}
{Hildebrandt}, H., {Viola}, M., {Heymans}, C., {et~al.} 2017, \mnras, 465, 1454

\bibitem[{{Hoyle}(2016)}]{Hoyle2016A&C....16...34H}
{Hoyle}, B. 2016, Astronomy and Computing, 16, 34

\bibitem[{Huber(1964)}]{Huber10.1214/aoms/1177703732}
Huber, P.~J. 1964, The Annals of Mathematical Statistics, 35, 73

\bibitem[{{Ivezi{\'c}} {et~al.}(2019){Ivezi{\'c}}, {Kahn}, {Tyson}, {Abel},
  {Acosta}, {Allsman}, {Alonso}, {AlSayyad}, {Anderson}, {Andrew}, {Angel},
  {Angeli}, {Ansari}, {Antilogus}, {Araujo}, {Armstrong}, {Arndt}, {Astier},
  {Aubourg}, {Auza}, {Axelrod}, {Bard}, {Barr}, {Barrau}, {Bartlett}, {Bauer},
  {Bauman}, {Baumont}, {Bechtol}, {Bechtol}, {Becker}, {Becla}, {Beldica},
  {Bellavia}, {Bianco}, {Biswas}, {Blanc}, {Blazek}, {Blandford}, {Bloom},
  {Bogart}, {Bond}, {Booth}, {Borgland}, {Borne}, {Bosch}, {Boutigny},
  {Brackett}, {Bradshaw}, {Brandt}, {Brown}, {Bullock}, {Burchat}, {Burke},
  {Cagnoli}, {Calabrese}, {Callahan}, {Callen}, {Carlin}, {Carlson},
  {Chandrasekharan}, {Charles-Emerson}, {Chesley}, {Cheu}, {Chiang}, {Chiang},
  {Chirino}, {Chow}, {Ciardi}, {Claver}, {Cohen-Tanugi}, {Cockrum}, {Coles},
  {Connolly}, {Cook}, {Cooray}, {Covey}, {Cribbs}, {Cui}, {Cutri}, {Daly},
  {Daniel}, {Daruich}, {Daubard}, {Daues}, {Dawson}, {Delgado}, {Dellapenna},
  {de Peyster}, {de Val-Borro}, {Digel}, {Doherty}, {Dubois},
  {Dubois-Felsmann}, {Durech}, {Economou}, {Eifler}, {Eracleous}, {Emmons},
  {Fausti Neto}, {Ferguson}, {Figueroa}, {Fisher-Levine}, {Focke}, {Foss},
  {Frank}, {Freemon}, {Gangler}, {Gawiser}, {Geary}, {Gee}, {Geha}, {Gessner},
  {Gibson}, {Gilmore}, {Glanzman}, {Glick}, {Goldina}, {Goldstein}, {Goodenow},
  {Graham}, {Gressler}, {Gris}, {Guy}, {Guyonnet}, {Haller}, {Harris},
  {Hascall}, {Haupt}, {Hernandez}, {Herrmann}, {Hileman}, {Hoblitt}, {Hodgson},
  {Hogan}, {Howard}, {Huang}, {Huffer}, {Ingraham}, {Innes}, {Jacoby}, {Jain},
  {Jammes}, {Jee}, {Jenness}, {Jernigan}, {Jevremovi{\'c}}, {Johns}, {Johnson},
  {Johnson}, {Jones}, {Juramy-Gilles}, {Juri{\'c}}, {Kalirai}, {Kallivayalil},
  {Kalmbach}, {Kantor}, {Karst}, {Kasliwal}, {Kelly}, {Kessler}, {Kinnison},
  {Kirkby}, {Knox}, {Kotov}, {Krabbendam}, {Krughoff}, {Kub{\'a}nek},
  {Kuczewski}, {Kulkarni}, {Ku}, {Kurita}, {Lage}, {Lambert}, {Lange},
  {Langton}, {Le Guillou}, {Levine}, {Liang}, {Lim}, {Lintott}, {Long},
  {Lopez}, {Lotz}, {Lupton}, {Lust}, {MacArthur}, {Mahabal}, {Mandelbaum},
  {Markiewicz}, {Marsh}, {Marshall}, {Marshall}, {May}, {McKercher}, {McQueen},
  {Meyers}, {Migliore}, {Miller}, {Mills}, {Miraval}, {Moeyens}, {Moolekamp},
  {Monet}, {Moniez}, {Monkewitz}, {Montgomery}, {Morrison}, {Mueller},
  {Muller}, {Mu{\~n}oz Arancibia}, {Neill}, {Newbry}, {Nief}, {Nomerotski},
  {Nordby}, {O'Connor}, {Oliver}, {Olivier}, {Olsen}, {O'Mullane}, {Ortiz},
  {Osier}, {Owen}, {Pain}, {Palecek}, {Parejko}, {Parsons}, {Pease},
  {Peterson}, {Peterson}, {Petravick}, {Libby Petrick}, {Petry},
  {Pierfederici}, {Pietrowicz}, {Pike}, {Pinto}, {Plante}, {Plate}, {Plutchak},
  {Price}, {Prouza}, {Radeka}, {Rajagopal}, {Rasmussen}, {Regnault}, {Reil},
  {Reiss}, {Reuter}, {Ridgway}, {Riot}, {Ritz}, {Robinson}, {Roby}, {Roodman},
  {Rosing}, {Roucelle}, {Rumore}, {Russo}, {Saha}, {Sassolas}, {Schalk},
  {Schellart}, {Schindler}, {Schmidt}, {Schneider}, {Schneider}, {Schoening},
  {Schumacher}, {Schwamb}, {Sebag}, {Selvy}, {Sembroski}, {Seppala}, {Serio},
  {Serrano}, {Shaw}, {Shipsey}, {Sick}, {Silvestri}, {Slater}, {Smith},
  {Smith}, {Sobhani}, {Soldahl}, {Storrie-Lombardi}, {Stover}, {Strauss},
  {Street}, {Stubbs}, {Sullivan}, {Sweeney}, {Swinbank}, {Szalay}, {Takacs},
  {Tether}, {Thaler}, {Thayer}, {Thomas}, {Thornton}, {Thukral}, {Tice},
  {Trilling}, {Turri}, {Van Berg}, {Vanden Berk}, {Vetter}, {Virieux},
  {Vucina}, {Wahl}, {Walkowicz}, {Walsh}, {Walter}, {Wang}, {Wang}, {Warner},
  {Wiecha}, {Willman}, {Winters}, {Wittman}, {Wolff}, {Wood-Vasey}, {Wu},
  {Xin}, {Yoachim}, \& {Zhan}}]{Izevic+19_LSST}
{Ivezi{\'c}}, {\v{Z}}., {Kahn}, S.~M., {Tyson}, J.~A., {et~al.} 2019, \apj,
  873, 111

\bibitem[{{Joachimi} {et~al.}(2021){Joachimi}, {Lin}, {Asgari}, {Tr{\"o}ster},
  {Heymans}, {Hildebrandt}, {K{\"o}hlinger}, {S{\'a}nchez}, {Wright},
  {Bilicki}, {Blake}, {van den Busch}, {Crocce}, {Dvornik}, {Erben}, {Getman},
  {Giblin}, {Hoekstra}, {Kannawadi}, {Kuijken}, {Napolitano}, {Schneider},
  {Scoccimarro}, {Sellentin}, {Shan}, {von Wietersheim-Kramsta}, \&
  {Zuntz}}]{Joachimi2021A&A...646A.129J}
{Joachimi}, B., {Lin}, C.~A., {Asgari}, M., {et~al.} 2021, \aap, 646, A129

\bibitem[{{Kendall} \& {Gal}(2017)}]{Kendall2017arXiv170304977K}
{Kendall}, A. \& {Gal}, Y. 2017, arXiv e-prints, arXiv:1703.04977

\bibitem[{{Khramtsov} {et~al.}(2019){Khramtsov}, {Sergeyev}, {Spiniello},
  {Tortora}, {Napolitano}, {Agnello}, {Getman}, {de Jong}, {Kuijken},
  {Radovich}, {Shan}, \& {Shulga}}]{Khramtsov2019A&A}
{Khramtsov}, V., {Sergeyev}, A., {Spiniello}, C., {et~al.} 2019, \aap, 632, A56

\bibitem[{{Kingma} \& {Ba}(2014)}]{Kingma2014+Adam}
{Kingma}, D.~P. \& {Ba}, J. 2014, arXiv e-prints, arXiv:1412.6980

\bibitem[{{Koo}(1985)}]{Koo1985AJ}
{Koo}, D.~C. 1985, \aj, 90, 418

\bibitem[{{Kuijken}(2011)}]{Kuijken2011Msngr.146....8K}
{Kuijken}, K. 2011, The Messenger, 146, 8

\bibitem[{{Kuijken} {et~al.}(2019){Kuijken}, {Heymans}, {Dvornik},
  {Hildebrandt}, {de Jong}, {Wright}, {Erben}, {Bilicki}, {Giblin}, {Shan},
  {Getman}, {Grado}, {Hoekstra}, {Miller}, {Napolitano}, {Paolilo}, {Radovich},
  {Schneider}, {Sutherland }, {Tewes}, {Tortora}, {Valentijn}, \& {Verdoes
  Kleijn}}]{Kuijken+19_KiDS-DR4}
{Kuijken}, K., {Heymans}, C., {Dvornik}, A., {et~al.} 2019, \aap, 625, A2

\bibitem[{{Kuijken} {et~al.}(2015){Kuijken}, {Heymans}, {Hildebrandt},
  {Nakajima}, {Erben}, {de Jong}, {Viola}, {Choi}, {Hoekstra}, {Miller}, {van
  Uitert}, {Amon}, {Blake}, {Brouwer}, {Buddendiek}, {Conti}, {Eriksen},
  {Grado}, {Harnois-D{\'e}raps}, {Helmich}, {Herbonnet}, {Irisarri},
  {Kitching}, {Klaes}, {La Barbera}, {Napolitano}, {Radovich}, {Schneider},
  {Sif{\'o}n}, {Sikkema}, {Simon}, {Tudorica}, {Valentijn}, {Verdoes Kleijn},
  \& {van Waerbeke}}]{Kuijken2015MNRAS}
{Kuijken}, K., {Heymans}, C., {Hildebrandt}, H., {et~al.} 2015, \mnras, 454,
  3500

\bibitem[{{Laureijs} {et~al.}(2011{\natexlab{a}}){Laureijs}, {Amiaux},
  {Arduini}, {Augu{\`e}res}, {Brinchmann}, {Cole}, {Cropper}, {Dabin}, {Duvet},
  {Ealet}, {Garilli}, {Gondoin}, {Guzzo}, {Hoar}, {Hoekstra}, {Holmes},
  {Kitching}, {Maciaszek}, {Mellier}, {Pasian}, {Percival}, {Rhodes}, {Saavedra
  Criado}, {Sauvage}, {Scaramella}, {Valenziano}, {Warren}, {Bender},
  {Castander}, {Cimatti}, {Le F{\`e}vre}, {Kurki-Suonio}, {Levi}, {Lilje},
  {Meylan}, {Nichol}, {Pedersen}, {Popa}, {Rebolo Lopez}, {Rix}, {Rottgering},
  {Zeilinger}, {Grupp}, {Hudelot}, {Massey}, {Meneghetti}, {Miller}, {Paltani},
  {Paulin-Henriksson}, {Pires}, {Saxton}, {Schrabback}, {Seidel}, {Walsh},
  {Aghanim}, {Amendola}, {Bartlett}, {Baccigalupi}, {Beaulieu}, {Benabed},
  {Cuby}, {Elbaz}, {Fosalba}, {Gavazzi}, {Helmi}, {Hook}, {Irwin}, {Kneib},
  {Kunz}, {Mannucci}, {Moscardini}, {Tao}, {Teyssier}, {Weller}, {Zamorani},
  {Zapatero Osorio}, {Boulade}, {Foumond}, {Di Giorgio}, {Guttridge}, {James},
  {Kemp}, {Martignac}, {Spencer}, {Walton}, {Bl{\"u}mchen}, {Bonoli},
  {Bortoletto}, {Cerna}, {Corcione}, {Fabron}, {Jahnke}, {Ligori}, {Madrid},
  {Martin}, {Morgante}, {Pamplona}, {Prieto}, {Riva}, {Toledo}, {Trifoglio},
  {Zerbi}, {Abdalla}, {Douspis}, {Grenet}, {Borgani}, {Bouwens}, {Courbin},
  {Delouis}, {Dubath}, {Fontana}, {Frailis}, {Grazian}, {Koppenh{\"o}fer},
  {Mansutti}, {Melchior}, {Mignoli}, {Mohr}, {Neissner}, {Noddle}, {Poncet},
  {Scodeggio}, {Serrano}, {Shane}, {Starck}, {Surace}, {Taylor},
  {Verdoes-Kleijn}, {Vuerli}, {Williams}, {Zacchei}, {Altieri}, {Escudero
  Sanz}, {Kohley}, {Oosterbroek}, {Astier}, {Bacon}, {Bardelli}, {Baugh},
  {Bellagamba}, {Benoist}, {Bianchi}, {Biviano}, {Branchini}, {Carbone},
  {Cardone}, {Clements}, {Colombi}, {Conselice}, {Cresci}, {Deacon}, {Dunlop},
  {Fedeli}, {Fontanot}, {Franzetti}, {Giocoli}, {Garcia-Bellido}, {Gow},
  {Heavens}, {Hewett}, {Heymans}, {Holland}, {Huang}, {Ilbert}, {Joachimi},
  {Jennins}, {Kerins}, {Kiessling}, {Kirk}, {Kotak}, {Krause}, {Lahav}, {van
  Leeuwen}, {Lesgourgues}, {Lombardi}, {Magliocchetti}, {Maguire}, {Majerotto},
  {Maoli}, {Marulli}, {Maurogordato}, {McCracken}, {McLure}, {Melchiorri},
  {Merson}, {Moresco}, {Nonino}, {Norberg}, {Peacock}, {Pello}, {Penny},
  {Pettorino}, {Di Porto}, {Pozzetti}, {Quercellini}, {Radovich}, {Rassat},
  {Roche}, {Ronayette}, {Rossetti}, {Sartoris}, {Schneider}, {Semboloni},
  {Serjeant}, {Simpson}, {Skordis}, {Smadja}, {Smartt}, {Spano}, {Spiro},
  {Sullivan}, {Tilquin}, {Trotta}, {Verde}, {Wang}, {Williger}, {Zhao},
  {Zoubian}, \& {Zucca}}]{Laureijs+11_Euclid}
{Laureijs}, R., {Amiaux}, J., {Arduini}, S., {et~al.} 2011{\natexlab{a}}, arXiv
  e-prints, arXiv:1110.3193

\bibitem[{{Laureijs} {et~al.}(2011{\natexlab{b}}){Laureijs}, {Amiaux},
  {Arduini}, {Augu{\`e}res}, {Brinchmann}, {Cole}, {Cropper}, {Dabin}, {Duvet},
  {Ealet}, {Garilli}, {Gondoin}, {Guzzo}, {Hoar}, {Hoekstra}, {Holmes},
  {Kitching}, {Maciaszek}, {Mellier}, {Pasian}, {Percival}, {Rhodes}, {Saavedra
  Criado}, {Sauvage}, {Scaramella}, {Valenziano}, {Warren}, {Bender},
  {Castander}, {Cimatti}, {Le F{\`e}vre}, {Kurki-Suonio}, {Levi}, {Lilje},
  {Meylan}, {Nichol}, {Pedersen}, {Popa}, {Rebolo Lopez}, {Rix}, {Rottgering},
  {Zeilinger}, {Grupp}, {Hudelot}, {Massey}, {Meneghetti}, {Miller}, {Paltani},
  {Paulin-Henriksson}, {Pires}, {Saxton}, {Schrabback}, {Seidel}, {Walsh},
  {Aghanim}, {Amendola}, {Bartlett}, {Baccigalupi}, {Beaulieu}, {Benabed},
  {Cuby}, {Elbaz}, {Fosalba}, {Gavazzi}, {Helmi}, {Hook}, {Irwin}, {Kneib},
  {Kunz}, {Mannucci}, {Moscardini}, {Tao}, {Teyssier}, {Weller}, {Zamorani},
  {Zapatero Osorio}, {Boulade}, {Foumond}, {Di Giorgio}, {Guttridge}, {James},
  {Kemp}, {Martignac}, {Spencer}, {Walton}, {Bl{\"u}mchen}, {Bonoli},
  {Bortoletto}, {Cerna}, {Corcione}, {Fabron}, {Jahnke}, {Ligori}, {Madrid},
  {Martin}, {Morgante}, {Pamplona}, {Prieto}, {Riva}, {Toledo}, {Trifoglio},
  {Zerbi}, {Abdalla}, {Douspis}, {Grenet}, {Borgani}, {Bouwens}, {Courbin},
  {Delouis}, {Dubath}, {Fontana}, {Frailis}, {Grazian}, {Koppenh{\"o}fer},
  {Mansutti}, {Melchior}, {Mignoli}, {Mohr}, {Neissner}, {Noddle}, {Poncet},
  {Scodeggio}, {Serrano}, {Shane}, {Starck}, {Surace}, {Taylor},
  {Verdoes-Kleijn}, {Vuerli}, {Williams}, {Zacchei}, {Altieri}, {Escudero
  Sanz}, {Kohley}, {Oosterbroek}, {Astier}, {Bacon}, {Bardelli}, {Baugh},
  {Bellagamba}, {Benoist}, {Bianchi}, {Biviano}, {Branchini}, {Carbone},
  {Cardone}, {Clements}, {Colombi}, {Conselice}, {Cresci}, {Deacon}, {Dunlop},
  {Fedeli}, {Fontanot}, {Franzetti}, {Giocoli}, {Garcia-Bellido}, {Gow},
  {Heavens}, {Hewett}, {Heymans}, {Holland}, {Huang}, {Ilbert}, {Joachimi},
  {Jennins}, {Kerins}, {Kiessling}, {Kirk}, {Kotak}, {Krause}, {Lahav}, {van
  Leeuwen}, {Lesgourgues}, {Lombardi}, {Magliocchetti}, {Maguire}, {Majerotto},
  {Maoli}, {Marulli}, {Maurogordato}, {McCracken}, {McLure}, {Melchiorri},
  {Merson}, {Moresco}, {Nonino}, {Norberg}, {Peacock}, {Pello}, {Penny},
  {Pettorino}, {Di Porto}, {Pozzetti}, {Quercellini}, {Radovich}, {Rassat},
  {Roche}, {Ronayette}, {Rossetti}, {Sartoris}, {Schneider}, {Semboloni},
  {Serjeant}, {Simpson}, {Skordis}, {Smadja}, {Smartt}, {Spano}, {Spiro},
  {Sullivan}, {Tilquin}, {Trotta}, {Verde}, {Wang}, {Williger}, {Zhao},
  {Zoubian}, \& {Zucca}}]{Laureijs2011arXiv1110.3193L}
{Laureijs}, R., {Amiaux}, J., {Arduini}, S., {et~al.} 2011{\natexlab{b}}, arXiv
  e-prints, arXiv:1110.3193

\bibitem[{{Li} {et~al.}(2021{\natexlab{a}}){Li}, {Napolitano}, {Roy},
  {Tortora}, {La Barbera}, {Sonnenfeld}, {Qiu}, \& {Liu}}]{Li2021_GaLNet}
{Li}, R., {Napolitano}, N.~R., {Roy}, N., {et~al.} 2021{\natexlab{a}}, arXiv
  e-prints, arXiv:2111.05434

\bibitem[{{Li} {et~al.}(2021{\natexlab{b}}){Li}, {Napolitano}, {Spiniello},
  {Tortora}, {Kuijken}, {Koopmans}, {Schneider}, {Getman}, {Xie}, {Long},
  {Shu}, {Vernardos}, {Huang}, {Covone}, {Dvornik}, {Heymans}, {Hildebrandt},
  {Radovich}, \& {Wright}}]{Li2021_DR5lens}
{Li}, R., {Napolitano}, N.~R., {Spiniello}, C., {et~al.} 2021{\natexlab{b}},
  \apj, 923, 16

\bibitem[{{Li} {et~al.}(2020){Li}, {Napolitano}, {Tortora}, {Spiniello},
  {Koopmans}, {Huang}, {Roy}, {Vernardos}, {Chatterjee}, {Giblin}, {Getman},
  {Radovich}, {Covone}, \& {Kuijken}}]{Li2020_DR4lens}
{Li}, R., {Napolitano}, N.~R., {Tortora}, C., {et~al.} 2020, \apj, 899, 30

\bibitem[{{Lilly} {et~al.}(2007){Lilly}, {Le F{\`e}vre}, {Renzini}, {Zamorani},
  {Scodeggio}, {Contini}, {Carollo}, {Hasinger}, {Kneib}, {Iovino}, {Le Brun},
  {Maier}, {Mainieri}, {Mignoli}, {Silverman}, {Tasca}, {Bolzonella},
  {Bongiorno}, {Bottini}, {Capak}, {Caputi}, {Cimatti}, {Cucciati}, {Daddi},
  {Feldmann}, {Franzetti}, {Garilli}, {Guzzo}, {Ilbert}, {Kampczyk}, {Kovac},
  {Lamareille}, {Leauthaud}, {Le Borgne}, {McCracken}, {Marinoni}, {Pello},
  {Ricciardelli}, {Scarlata}, {Vergani}, {Sanders}, {Schinnerer}, {Scoville},
  {Taniguchi}, {Arnouts}, {Aussel}, {Bardelli}, {Brusa}, {Cappi}, {Ciliegi},
  {Finoguenov}, {Foucaud}, {Franceschini}, {Halliday}, {Impey}, {Knobel},
  {Koekemoer}, {Kurk}, {Maccagni}, {Maddox}, {Marano}, {Marconi}, {Meneux},
  {Mobasher}, {Moreau}, {Peacock}, {Porciani}, {Pozzetti}, {Scaramella},
  {Schiminovich}, {Shopbell}, {Smail}, {Thompson}, {Tresse}, {Vettolani},
  {Zanichelli}, \& {Zucca}}]{Lilly2007ApJS_zcosmos}
{Lilly}, S.~J., {Le F{\`e}vre}, O., {Renzini}, A., {et~al.} 2007, \apjs, 172,
  70

\bibitem[{{LSST Science Collaboration} {et~al.}(2009){LSST Science
  Collaboration}, {Abell}, {Allison}, {Anderson}, {Andrew}, {Angel}, {Armus},
  {Arnett}, {Asztalos}, {Axelrod}, {Bailey}, {Ballantyne}, {Bankert},
  {Barkhouse}, {Barr}, {Barrientos}, {Barth}, {Bartlett}, {Becker}, {Becla},
  {Beers}, {Bernstein}, {Biswas}, {Blanton}, {Bloom}, {Bochanski}, {Boeshaar},
  {Borne}, {Bradac}, {Brandt}, {Bridge}, {Brown}, {Brunner}, {Bullock},
  {Burgasser}, {Burge}, {Burke}, {Cargile}, {Chandrasekharan}, {Chartas},
  {Chesley}, {Chu}, {Cinabro}, {Claire}, {Claver}, {Clowe}, {Connolly}, {Cook},
  {Cooke}, {Cooray}, {Covey}, {Culliton}, {de Jong}, {de Vries}, {Debattista},
  {Delgado}, {Dell'Antonio}, {Dhital}, {Di Stefano}, {Dickinson}, {Dilday},
  {Djorgovski}, {Dobler}, {Donalek}, {Dubois-Felsmann}, {Durech},
  {Eliasdottir}, {Eracleous}, {Eyer}, {Falco}, {Fan}, {Fassnacht}, {Ferguson},
  {Fernandez}, {Fields}, {Finkbeiner}, {Figueroa}, {Fox}, {Francke}, {Frank},
  {Frieman}, {Fromenteau}, {Furqan}, {Galaz}, {Gal-Yam}, {Garnavich},
  {Gawiser}, {Geary}, {Gee}, {Gibson}, {Gilmore}, {Grace}, {Green}, {Gressler},
  {Grillmair}, {Habib}, {Haggerty}, {Hamuy}, {Harris}, {Hawley}, {Heavens},
  {Hebb}, {Henry}, {Hileman}, {Hilton}, {Hoadley}, {Holberg}, {Holman},
  {Howell}, {Infante}, {Ivezic}, {Jacoby}, {Jain}, {R}, {Jedicke}, {Jee},
  {Garrett Jernigan}, {Jha}, {Johnston}, {Jones}, {Juric}, {Kaasalainen},
  {Styliani}, {Kafka}, {Kahn}, {Kaib}, {Kalirai}, {Kantor}, {Kasliwal},
  {Keeton}, {Kessler}, {Knezevic}, {Kowalski}, {Krabbendam}, {Krughoff},
  {Kulkarni}, {Kuhlman}, {Lacy}, {Lepine}, {Liang}, {Lien}, {Lira}, {Long},
  {Lorenz}, {Lotz}, {Lupton}, {Lutz}, {Macri}, {Mahabal}, {Mandelbaum},
  {Marshall}, {May}, {McGehee}, {Meadows}, {Meert}, {Milani}, {Miller},
  {Miller}, {Mills}, {Minniti}, {Monet}, {Mukadam}, {Nakar}, {Neill}, {Newman},
  {Nikolaev}, {Nordby}, {O'Connor}, {Oguri}, {Oliver}, {Olivier}, {Olsen},
  {Olsen}, {Olszewski}, {Oluseyi}, {Padilla}, {Parker}, {Pepper}, {Peterson},
  {Petry}, {Pinto}, {Pizagno}, {Popescu}, {Prsa}, {Radcka}, {Raddick},
  {Rasmussen}, {Rau}, {Rho}, {Rhoads}, {Richards}, {Ridgway}, {Robertson},
  {Roskar}, {Saha}, {Sarajedini}, {Scannapieco}, {Schalk}, {Schindler},
  {Schmidt}, {Schmidt}, {Schneider}, {Schumacher}, {Scranton}, {Sebag},
  {Seppala}, {Shemmer}, {Simon}, {Sivertz}, {Smith}, {Allyn Smith}, {Smith},
  {Spitz}, {Stanford}, {Stassun}, {Strader}, {Strauss}, {Stubbs}, {Sweeney},
  {Szalay}, {Szkody}, {Takada}, {Thorman}, {Trilling}, {Trimble}, {Tyson}, {Van
  Berg}, {Vanden Berk}, {VanderPlas}, {Verde}, {Vrsnak}, {Walkowicz},
  {Wandelt}, {Wang}, {Wang}, {Warner}, {Wechsler}, {West}, {Wiecha},
  {Williams}, {Willman}, {Wittman}, {Wolff}, {Wood-Vasey}, {Wozniak}, {Young},
  {Zentner}, \& {Zhan}}]{LSST2009arXiv0912.0201L}
{LSST Science Collaboration}, {Abell}, P.~A., {Allison}, J., {et~al.} 2009,
  arXiv e-prints, arXiv:0912.0201

\bibitem[{{Moster} {et~al.}(2013){Moster}, {Naab}, \&
  {White}}]{Moster2013MNRAS.428.3121M}
{Moster}, B.~P., {Naab}, T., \& {White}, S. D.~M. 2013, \mnras, 428, 3121

\bibitem[{{Nakoneczny} {et~al.}(2019){Nakoneczny}, {Bilicki}, {Solarz},
  {Pollo}, {Maddox}, {Spiniello}, {Brescia}, \&
  {Napolitano}}]{2019A&A...624A..13N}
{Nakoneczny}, S., {Bilicki}, M., {Solarz}, A., {et~al.} 2019, \aap, 624, A13

\bibitem[{{Newman} {et~al.}(2013){Newman}, {Cooper}, {Davis}, {Faber}, {Coil},
  {Guhathakurta}, {Koo}, {Phillips}, {Conroy}, {Dutton}, {Finkbeiner}, {Gerke},
  {Rosario}, {Weiner}, {Willmer}, {Yan}, {Harker}, {Kassin}, {Konidaris},
  {Lai}, {Madgwick}, {Noeske}, {Wirth}, {Connolly}, {Kaiser}, {Kirby},
  {Lemaux}, {Lin}, {Lotz}, {Luppino}, {Marinoni}, {Matthews}, {Metevier}, \&
  {Schiavon}}]{Newman2013ApJS_DEEP2}
{Newman}, J.~A., {Cooper}, M.~C., {Davis}, M., {et~al.} 2013, \apjs, 208, 5

\bibitem[{{Pasquet} {et~al.}(2019){Pasquet}, {Bertin}, {Treyer}, {Arnouts}, \&
  {Fouchez}}]{Pasquet2019A&A...621A..26P}
{Pasquet}, J., {Bertin}, E., {Treyer}, M., {Arnouts}, S., \& {Fouchez}, D.
  2019, \aap, 621, A26

\bibitem[{{Podsztavek} {et~al.}(2022){Podsztavek}, {{\v{S}}koda}, \&
  {Tvrd{\'\i}k}}]{Podsztavek2022ascl.soft04004P}
{Podsztavek}, O., {{\v{S}}koda}, P., \& {Tvrd{\'\i}k}, P. 2022, {Bayesian
  SZNet: Bayesian deep learning to predict redshift with uncertainty},
  Astrophysics Source Code Library, record ascl:2204.004

\bibitem[{{Ramachandra} {et~al.}(2021){Ramachandra}, {Chaves-Montero},
  {Alarcon}, {Fadikar}, {Habib}, \&
  {Heitmann}}]{Ramachandra2021arXiv211112118R}
{Ramachandra}, N., {Chaves-Montero}, J., {Alarcon}, A., {et~al.} 2021, arXiv
  e-prints, arXiv:2111.12118

\bibitem[{{Rhea} {et~al.}(2021){Rhea}, {Hlavacek-Larrondo}, {Rousseau-Nepton},
  \& {Prunet}}]{Rhea2021RNAAS...5..276R}
{Rhea}, C., {Hlavacek-Larrondo}, J., {Rousseau-Nepton}, L., \& {Prunet}, S.
  2021, Research Notes of the American Astronomical Society, 5, 276

\bibitem[{{Roy} {et~al.}(2018){Roy}, {Napolitano}, {La Barbera}, {Tortora},
  {Getman}, {Radovich}, {Capaccioli}, {Brescia}, {Cavuoti}, {Longo}, {Raj},
  {Puddu}, {Covone}, {Amaro}, {Vellucci}, {Grado}, {Kuijken}, {Verdoes Kleijn},
  \& {Valentijn}}]{Roy+18}
{Roy}, N., {Napolitano}, N.~R., {La Barbera}, F., {et~al.} 2018, \mnras, 480,
  1057

\bibitem[{{Sadeh} {et~al.}(2016){Sadeh}, {Abdalla}, \&
  {Lahav}}]{Sadeh2016PASP..128j4502S}
{Sadeh}, I., {Abdalla}, F.~B., \& {Lahav}, O. 2016, \pasp, 128, 104502

\bibitem[{{Schmidt} {et~al.}(2020){Schmidt}, {Malz}, {Soo}, {Almosallam},
  {Brescia}, {Cavuoti}, {Cohen-Tanugi}, {Connolly}, {DeRose}, {Freeman},
  {Graham}, {Iyer}, {Jarvis}, {Kalmbach}, {Kovacs}, {Lee}, {Longo}, {Morrison},
  {Newman}, {Nourbakhsh}, {Nuss}, {Pospisil}, {Tranin}, {Wechsler}, {Zhou},
  {Izbicki}, \& {LSST Dark Energy Science
  Collaboration}}]{Schmidt2020MNRAS.499.1587S}
{Schmidt}, S.~J., {Malz}, A.~I., {Soo}, J.~Y.~H., {et~al.} 2020, \mnras, 499,
  1587

\bibitem[{{Schuldt} {et~al.}(2021){Schuldt}, {Suyu}, {Ca{\~n}ameras},
  {Taubenberger}, {Meinhardt}, {Leal-Taix{\'e}}, \&
  {Hsieh}}]{Schuldt2021A&A...651A..55S}
{Schuldt}, S., {Suyu}, S.~H., {Ca{\~n}ameras}, R., {et~al.} 2021, \aap, 651,
  A55

\bibitem[{{Simonyan} \& {Zisserman}(2014)}]{Simonyan2014}
{Simonyan}, K. \& {Zisserman}, A. 2014, arXiv e-prints, arXiv:1409.1556

\bibitem[{{Soo} {et~al.}(2018){Soo}, {Moraes}, {Joachimi}, {Hartley}, {Lahav},
  {Charbonnier}, {Makler}, {Pereira}, {Comparat}, {Erben}, {Leauthaud}, {Shan},
  \& {Van Waerbeke}}]{Soo2018MNRAS.475.3613S}
{Soo}, J. Y.~H., {Moraes}, B., {Joachimi}, B., {et~al.} 2018, \mnras, 475, 3613

\bibitem[{{Sutherland} {et~al.}(2015){Sutherland}, {Emerson}, {Dalton},
  {Atad-Ettedgui}, {Beard}, {Bennett}, {Bezawada}, {Born}, {Caldwell}, {Clark},
  {Craig}, {Henry}, {Jeffers}, {Little}, {McPherson}, {Murray}, {Stewart},
  {Stobie}, {Terrett}, {Ward}, {Whalley}, \&
  {Woodhouse}}]{Sutherland2015A&A...575A..25S_VISTA}
{Sutherland}, W., {Emerson}, J., {Dalton}, G., {et~al.} 2015, \aap, 575, A25

\bibitem[{{Szokoly} {et~al.}(2004){Szokoly}, {Bergeron}, {Hasinger}, {Lehmann},
  {Kewley}, {Mainieri}, {Nonino}, {Rosati}, {Giacconi}, {Gilli}, {Gilmozzi},
  {Norman}, {Romaniello}, {Schreier}, {Tozzi}, {Wang}, {Zheng}, \&
  {Zirm}}]{Szokoly2004ApJS_CDFS}
{Szokoly}, G.~P., {Bergeron}, J., {Hasinger}, G., {et~al.} 2004, \apjs, 155,
  271

\bibitem[{{The Dark Energy Survey Collaboration}(2005)}]{2005+DES}
{The Dark Energy Survey Collaboration}. 2005, arXiv e-prints, astro

\bibitem[{{Tohill} {et~al.}(2021){Tohill}, {Ferreira}, {Conselice}, {Bamford},
  \& {Ferrari}}]{2021ApJ...916....4T}
{Tohill}, C., {Ferreira}, L., {Conselice}, C.~J., {Bamford}, S.~P., \&
  {Ferrari}, F. 2021, \apj, 916, 4

\bibitem[{{Tortora} \& {Napolitano}(2022)}]{2022FrASS...8..197N}
{Tortora}, C. \& {Napolitano}, N.~R. 2022, Frontiers in Astronomy and Space
  Sciences, 8, 197

\bibitem[{{van den Busch} {et~al.}(2020){van den Busch}, {Hildebrandt},
  {Wright}, {Morrison}, {Blake}, {Joachimi}, {Erben}, {Heymans}, {Kuijken}, \&
  {Taylor}}]{Busch2020A&A...642A.200V}
{van den Busch}, J.~L., {Hildebrandt}, H., {Wright}, A.~H., {et~al.} 2020,
  \aap, 642, A200

\bibitem[{{Walmsley} {et~al.}(2020){Walmsley}, {Smith}, {Lintott}, {Gal},
  {Bamford}, {Dickinson}, {Fortson}, {Kruk}, {Masters}, {Scarlata}, {Simmons},
  {Smethurst}, \& {Wright}}]{2020MNRAS.491.1554W}
{Walmsley}, M., {Smith}, L., {Lintott}, C., {et~al.} 2020, \mnras, 491, 1554

\bibitem[{{Wang} {et~al.}(2022){Wang}, {Cheng}, {Ma}, \&
  {Xia}}]{Wang2022arXiv220700185W}
{Wang}, G.-J., {Cheng}, C., {Ma}, Y.-Z., \& {Xia}, J.-Q. 2022, arXiv e-prints,
  arXiv:2207.00185

\bibitem[{{Wang} {et~al.}(1998){Wang}, {Bahcall}, \& {Turner}}]{Wang1998AJ}
{Wang}, Y., {Bahcall}, N., \& {Turner}, E.~L. 1998, \aj, 116, 2081

\bibitem[{{Weinberg} {et~al.}(2013){Weinberg}, {Mortonson}, {Eisenstein},
  {Hirata}, {Riess}, \& {Rozo}}]{Weinberg2013PhR...530...87W}
{Weinberg}, D.~H., {Mortonson}, M.~J., {Eisenstein}, D.~J., {et~al.} 2013,
  \physrep, 530, 87

\bibitem[{{Yang} {et~al.}(2012){Yang}, {Mo}, {van den Bosch}, {Zhang}, \&
  {Han}}]{Yang2012ApJ...752...41Y}
{Yang}, X., {Mo}, H.~J., {van den Bosch}, F.~C., {Zhang}, Y., \& {Han}, J.
  2012, \apj, 752, 41

\bibitem[{{Zhan}(2018)}]{Zhan+18_csst}
{Zhan}, H. 2018, in 42nd COSPAR Scientific Assembly, Vol.~42, E1.16--4--18

\bibitem[{{Zhou} {et~al.}(2021){Zhou}, {Gong}, {Meng}, {Cao}, {Chen}, {Chen},
  {Du}, {Fu}, \& {Luo}}]{Zhou2021arXiv211208690Z}
{Zhou}, X., {Gong}, Y., {Meng}, X.-M., {et~al.} 2021, arXiv e-prints,
  arXiv:2112.08690

\end{thebibliography}

\end{document}